\documentclass[final,5p,times,twocolumn]{elsarticle}

\usepackage{graphicx}
\usepackage{subfigure}
\usepackage{multirow}
\usepackage{wrapfig}
\usepackage{array}

\usepackage{tabularray}
\usepackage{tabularx}
\usepackage{rotating}
\usepackage{makecell}
\usepackage{booktabs}
\usepackage{csquotes}

\usepackage{pifont}
\usepackage{verbatim}
\usepackage{multirow}
\usepackage{gensymb}
\usepackage{algorithm,algorithmic}
\usepackage{afterpage}
\usepackage{float}
\usepackage{lipsum}

\usepackage[utf8]{inputenc}
\usepackage{amsmath}
\usepackage{amsthm}
\usepackage{amsfonts}
\usepackage{epsfig}
\usepackage{psfrag}
\usepackage{pstricks}
\usepackage{algorithm}
\usepackage{stfloats}

\usepackage{cuted}

\graphicspath{{./Figures/}}

\usepackage{amssymb}
\usepackage{bm}
\usepackage{hyperref}
\hypersetup{
            colorlinks=true,
            linkcolor=blue,
            anchorcolor=blue,
            citecolor=blue
            }

\biboptions{comma,sort&compress}

\usepackage{enumitem}
\setlist{nosep,topsep=-\parskip}

\usepackage{ulem}

\journal{Elsevier}

\begin{document}

\begin{frontmatter}

\title{A Quasi-Optimal Shape Design Method for Lattice Structure Construction}
\author[zju]{Sifan Chen}
\author[sjtu]{Yuan Kong}
\author[zju]{Qiang Zou\corref{cor}}\ead{qiangzou@cad.zju.edu.cn}

\cortext[cor]{Corresponding author.}
\address[zju]{State Key Laboratory of CAD$\&$CG, Zhejiang University, Hangzhou, 310058, China}
\address[sjtu]{School of Materials Sciences and Engineering, Shanghai JiaoTong University, Shanghai 200240, China}
   
\begin{abstract}
Lattice structures, known for their superior mechanical properties, are widely used in industries such as aerospace, automotive, and biomedical. Their advantages primarily lie in the interconnected struts at the micro-scale. The robust construction of these struts is crucial for downstream design and manufacturing applications, as it provides a detailed shape description necessary for precise simulation and fabrication. However, constructing lattice structures presents significant challenges, particularly at nodes where multiple struts intersect. The complexity of these intersections can lead to robustness issues. To address this challenge, this paper presents an optimization-based approach that simplifies the construction of lattice structures by cutting struts and connecting them to optimized node shapes. By utilizing the recent Grey Wolf optimization method---a type of meta-heuristic method---for node shape design, the approach ensures robust model construction and optimal shape design. Its effectiveness has been validated through a series of case studies with increasing topological and geometric complexity.
\end{abstract}

\end{frontmatter}

\section{Introduction}
\label{sec:intro}

Lattice structures, composed of interconnected struts at the micro-scale, have found increasing applications in industries such as aerospace, automotive, and biomedical fields due to their superior mechanical properties and lightweight nature~\cite{zou2024geometric}. These structures enable efficient material use while maintaining structural integrity, making them ideal for performance- and weight-critical applications. Additionally, advancements in additive manufacturing technologies have made the production of these complex structures increasingly feasible~\cite{liu2021survey,nazir2019state}.

Extensive research has been carried out to understand the mechanical behavior of lattice structures and report the ``shape-property-process" relationship, both theoretically and empirically~\cite{wu2019design}. However, as recently noted by Verma et al.~\cite{verma2020combinatorial} and others~\cite{tao2016design,wang2005hybrid}, researchers have consistently encountered challenges in the construction of lattice structures while developing this relationship. Traditional solid modeling techniques in computer-aided design (CAD)~\cite{zou2023variational} often fail to address these challenges effectively~\cite{zou2024meta}.

\begin{figure}[t]
\centering
\includegraphics[width=\linewidth]{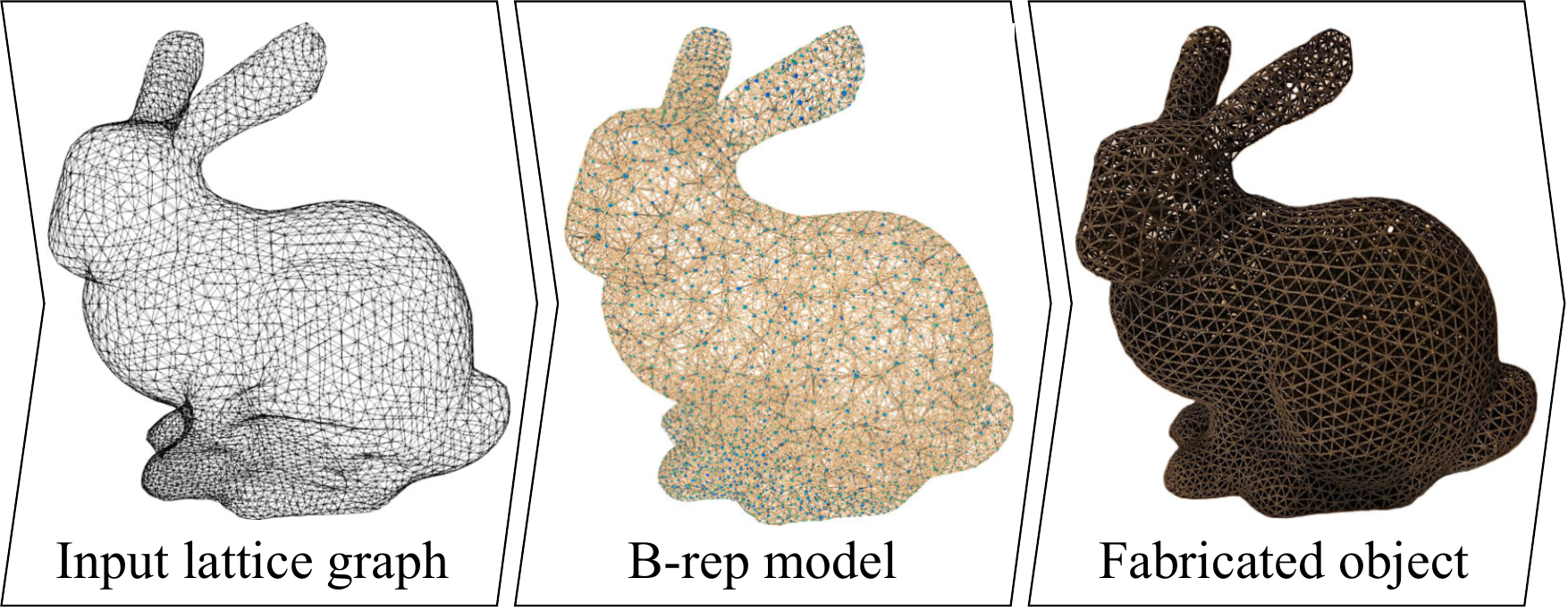}
\caption{Illustration of the lattice graph representation, the B-rep model constructed, and the final fabricated object} 
\label{fig:fabricate} 
\end{figure}

\begin{figure}[t]
\centering
\includegraphics[width=\linewidth]{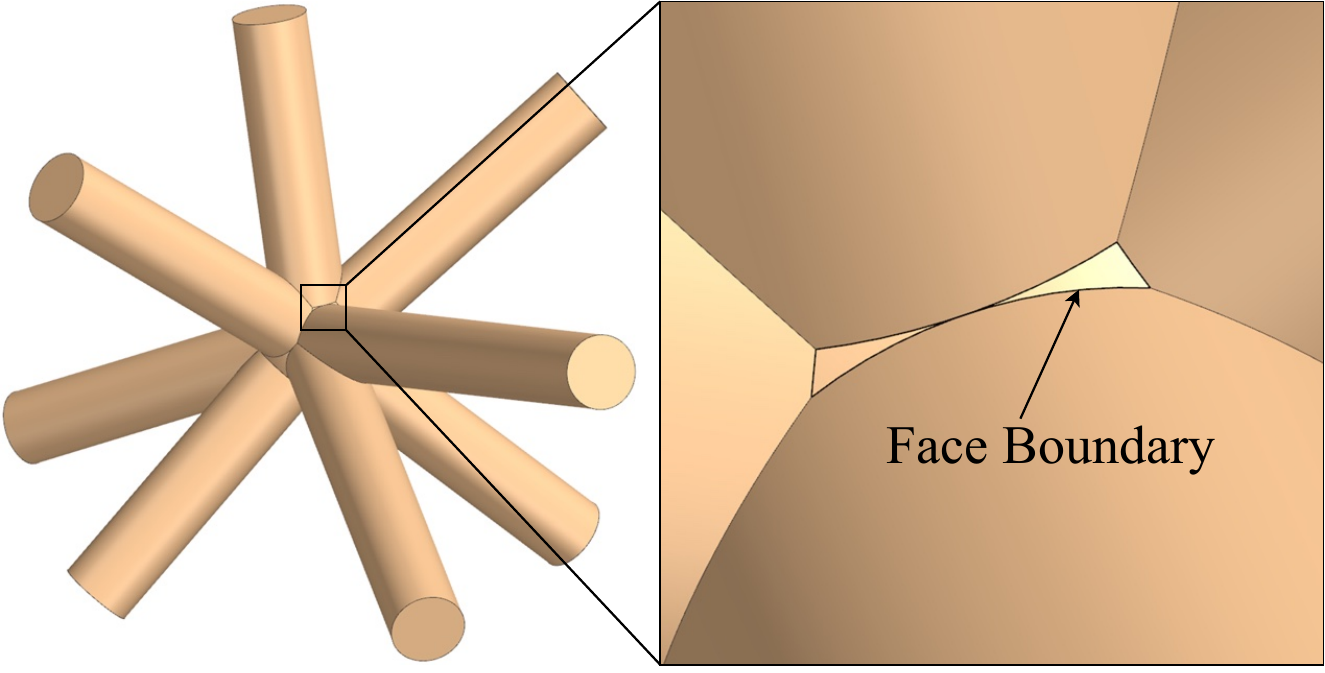}
\caption{A singular example (tangency) of boundary evaluation for a lattice node} 
\label{fig:robust} 
\end{figure}

To bridge this gap, a simple and robust method for lattice structure construction is developed in this work. Specifically, given a weighted lattice graph---comprising edges, nodes, and weights that define the struts, connections, and parameters (e.g., radii)---the goal is to create a watertight boundary representation (B-rep) model of the lattice structure for use in applications such as mesh generation~\cite{wang2018hex,chen2018finite}, finite element analysis~\cite{zou2023xvoxel,gu2021material}, and tool path generation~\cite{zou2013iso,zou2014iso}, as shown in Fig.~\ref{fig:fabricate}. A key challenge in this process is ensuring robustness in generating nodal shapes, as nodes represent intersections of multiple struts and involve complex Boolean operations to evaluate their boundary shapes. These operations are notoriously unreliable~\cite{verma2020combinatorial,liu2021memory}. Although struts (i.e., cylinders) are seemingly simple to intersect, robustness issues still occur. 
As already highlighted by Hoffman~\cite{hoffmann1989problems}, even for such simple geometries, singular tangencies and small face fragments (see Fig.~\ref{fig:robust}) can cause intersection failures.

To address these robustness challenges, two main strategies have been developed. The first, implicit modeling, involves converting the lattice structure into implicit fields, performing Boolean operations on these fields, and then converting the results to meshes via iso-contouring~\cite{pasko2011procedural,fryazinov2013multi,ma2022novel}. While this approach is robust, it is computationally intensive and tends to generate large numbers of unnecessary or distorted triangles. The second strategy simplifies the nodal shapes, typically using convex hulls, to avoid Boolean operations. While promising results have been demonstrated~\cite{verma2020combinatorial,wu2020chocc}, current methods do not address the deviation between simplified and original nodal shapes, nor do they determine which portions of the nodal shape are best to be replaced.

In this paper, we adopt the second strategy but enhance it with an optimization-based approach to reduce shape deviation. Specifically, we determine the optimal locations on each strut to cut the node from its adjacent struts, effectively separating the replaceable portion of the lattice from the parts to be retained. The goal is to preserve the original shape as much as possible. After the cuts, we apply a shape optimization method to fill the void left by the removed nodes. Given the high nonlinearity of this optimization problem, we use the Grey Wolf Optimizer (GWO)~\cite{mirjalili2014grey} to solve it, a meta-heuristic algorithm known for its efficiency due to simple solution update schemes and high parallelizability. This optimizer generates high-quality nodal shapes that are watertight, smooth, and self-intersection-free, and closely match the original nodal shapes. To the best of our knowledge, while meta-heuristic optimization has been widely used in engineering, this seems to be the first work of applying such techniques to lattice structure construction.

The remainder of this paper is organized as follows: Sec.~\ref{sec:related_work} reviews related studies, Sec.~\ref{sec:methods} details the proposed method, Sec.~\ref{sec:results} presents validation results through case studies, and Sec.~\ref{sec:conclusion} concludes with a discussion of the method’s advantages and limitations.

\section{Related Work}
\label{sec:related_work}
Existing lattice structure construction methods may be classified into two general categories: explicit and implicit. The former directly manipulates B-rep elements, while the latter constructs lattice structures using implicit fields. These methods are discussed in detail below.

\subsection{Explicit Methods}
The conceptually straightforward approach to constructing lattice structures involves modeling struts as cylindrical solids and nodes as spherical solids, which are then combined using Boolean operations to complete the structure. The efficiency, accuracy, and robustness of this method largely depend on the representation schemes used to define these cylindrical and spherical solids. A variety of modeling techniques have been explored, including algebraic surface-based models~\cite{wang2005hybrid,dumas2017modelling}, parametric surface-based models~\cite{2021_spherePacking_conformal}, subdivision surface-based models~\cite{2018_subdivision_2Dsurface,2019_differentialOffsetGrading,2023_2Dsurface_Xiong_Subdivisional-modelling}, and mesh-based models~\cite{chen20073d,2017_light-weight-triangulation,2017_prefab-cell_HGM}.

Algebraic and parametric surface models provide high accuracy due to their continuous nature, but their Boolean intersections can be computationally intensive and usually unreliable. Mesh-based models, by contrast, enable simpler intersection calculations; however, their accuracy and efficiency depend on the mesh resolution---higher resolution improves accuracy but reduces efficiency, and vice versa. Furthermore, ensuring robustness in mesh-based Boolean operations remains an ongoing challenge~\cite{trettner2022emberm}.

There are also studies that resort to the ``stitching" operation to construct B-reps of lattice structures~\cite{2017_light-weight-triangulation, verma2020combinatorial,wu2020chocc}. The idea is to generate triangular meshes for the struts and nodes separately, then stitch the vertices at their interfaces to create a watertight mesh for the entire lattice. These methods primarily differ in how the triangular meshes for the nodes are generated. In the work by Chougrani et al.~\cite{2017_light-weight-triangulation}, spherical nodes and strut-strut intersection curves are sampled, and the resulting points are connected to form a triangular mesh that approximates the nodal shape. (Notably, the intersection curves do not need to be pre-calculated; instead, sample points along the curves are directly obtained.) In contrast, Verma et al.~\cite{verma2020combinatorial} use convex hulls to approximate the nodal shape and stitch these hulls directly with the meshed, open struts. Similarly, Wu et al.~\cite{wu2020chocc} employ convex hulls of co-spherical circles to approximate the nodal shapes, resulting in improved shape quality, and this method is not restricted to triangular meshes. 

By shifting from solid Booleans to nodal shape approximations, complex strut-strut intersections are avoided, leading to improved robustness. While promising results have been demonstrated~\cite{verma2020combinatorial, wu2020chocc}, current methods still have room for improvement. Particularly, these approaches do not account for the potential deviations between approximate and original nodal shapes, nor do they provide a systematic method for identifying which portions of the nodal shape should be replaced with approximated versions. This paper aims to address this gap by proposing an optimization-based approach to reduce shape deviations.

\subsection{Implicit Methods}
Implicit modeling is a technique for representing and constructing 3D objects using mathematical functions and their iso-contours, rather than explicitly defining the geometry through B-reps. This approach naturally lends itself to Boolean operations, as they can be expressed algebraically by combining underlying implicit fields with max/min operations, which are straightforward to implement~\cite{1995_Pasko_F-rep}. Due to this simplicity, various implicit modeling methods have been applied to lattice modeling, where distance functions, convolutional surfaces, and even general functions (e.g., F-rep) are used to construct lattice structures~\cite{liu2021memory,ding2021stl,tang2019hybrid,pasko2011procedural,fryazinov2013multi,2024_F-rep_zhao,aremu2017voxel}. In recent years, software platforms such as nTopology~\cite{ntopology} have leveraged implicit modeling to enable more flexible construction of lattice structures. 

Despite its simplicity and robustness, the use of implicit modeling often requires sophisticated iso-contouring algorithms to extract the zero-value iso-surface from the combined implicit functions, converting implicit functions into explicit meshes~\cite{2020_subdivision-contour-extraction}. While such algorithms exist, such as marching cubes~\cite{newman2006survey} and dual-contouring~\cite{ju2002dual}, the quality of the resulting mesh surface depends heavily on the resolution of the underlying voxel grid, 
and redundant facets are often generated in this process. Additionally, many lattice structures cannot be effectively represented by analytical functions, necessitating the use of discrete implicit fields; for discrete fields, achieving high representation accuracy necessitates high resolution, which in turn results in high memory consumption. For these reasons, implicit modeling is most suitable for lattice structures that are analytically describable or for smaller-scale applications.

\section{Methods}
\label{sec:methods}

\subsection{Problem Statement}
\label{sec:problem}
We assume that a graph representation for the lattice structure of interest has been given. Specifically, the graph is defined as a tuple $ G = (V, E, W) $, where $ V $ is a set of vertices, $ E $ is a set of unordered pairs $ \{v_1, v_2\} \subseteq V $ representing edges, and $ W $ is a set of weights assigned to the nodes or edges. The vertices correspond to the nodes and their 3D coordinates, the edges represent struts connecting pairs of nodes, and the weights indicate the user-specified radius for each edge. (Note that radii can also be assigned to nodes; however, we adopt the edge-centric convention, which is more commonly used in engineering practice~\cite{verma2020combinatorial}.)

To store the weighted graph, we use the adjacency list data structure. This structure consists of a collection of lists, where each node points to a list of its adjacent nodes, along with the weight of the edge connecting them. The adjacency list is chosen for its space efficiency, particularly for sparsely connected graphs, compared to other representations like the adjacency matrix. This choice is well-suited to lattice structures, where each node is typically connected to only a small number of nearby nodes, rather than being linked to scattered nodes throughout the lattice.

Given the above input, the method to be presented outputs a B-rep model of the lattice structure with the following characteristics:
\begin{enumerate}
    \item The model is a valid solid model, a basic requirement for use in downstream design and manufacturing applications;
    \item The model consists of planar or curved faces (which are not necessarily triangles, in contrast to existing approaches~\cite{verma2020combinatorial, 2017_light-weight-triangulation}), enabling a more flexible and accurate lattice structure construction; and
    \item The model preserves the exact lattice shape as closely as possible.
\end{enumerate}
The term ``valid" in the first characteristic is evaluated to the following conditions~\cite{zou2019push}:
\begin{enumerate}
    \item A face is a bounded, regular~\footnote{See Ref.~\cite{requicha1980representations} for the definition of regular or regularization.} and semi-analytic subset of a piecewise analytic surface in $ \mathbb{R}^3 $, and the face interior must be connected;
    \item Each edge is shared by exactly two faces;
    \item Faces around each vertex form a closed fan; and
    \item Faces may intersect only at common edges or vertices.
\end{enumerate}
Condition 1 essentially requires faces to be well-bounded, meaning each face boundary is not open or intersected. Conditions 2 and 3 ensure manifoldness along each edge and around each vertex, while Condition 4 ensures manifoldness of the face interior.

\subsection{The Overall Workflow}
\label{sec:overview}
The fundamental principle of the proposed method is to replace the nodal shapes of a lattice structure with approximations, thereby avoiding the complex issue of surface-to-surface intersections. This approach is similar to the ideas presented by Verma et al.~\cite{verma2020combinatorial} and Wu et al.~\cite{wu2020chocc}, but it differs by considering the optimality of the replacement, which they do not address. Specifically, we carry out the replacement by a three-step process: 
\begin{enumerate}
    \item \textbf{Optimal cutting}: identifying the most suitable portions of a lattice structure's nodes to be cut off from its adjacent struts and replaced with approximations;
    \item \textbf{Optimal nodal shape design}: determining the best approximations for the cut portions, ensuring that the resulting shapes maintain a small deviation from the overall geometry of the structure; and
    \item \textbf{Lattice stitching}: integrating the approximated portions back into the lattice, ensuring a valid solid model for the given lattice structure.
\end{enumerate}
We also provide an additional step of multiresolution triangular mesh generation, which is useful for downstream applications such as visualization and slicing.

\begin{figure*}[t]
    \centering
    \includegraphics[width=\linewidth]{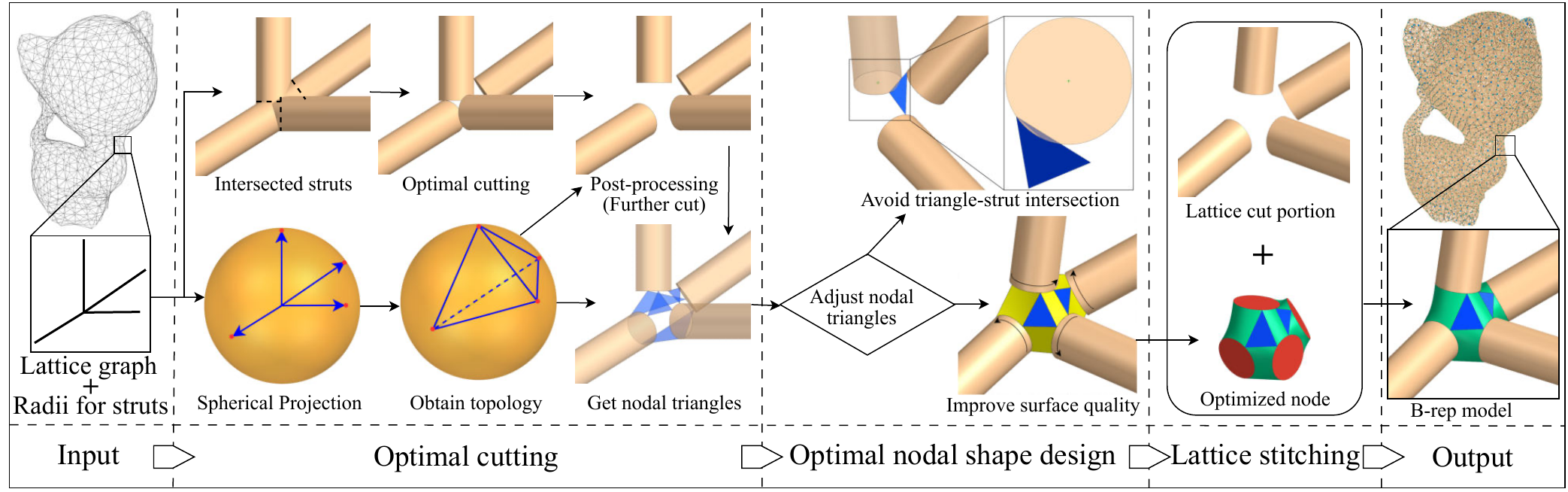}
    \caption{The proposed lattice structure construction workflow}
    \label{fig:pipeline} 
\end{figure*}

Fig.~\ref{fig:pipeline} illustrates the workflow and key points of the three steps described above. Step 1 takes a lattice graph (as described in Sec.~\ref{sec:problem}) as input and then determines an optimal cut location on each strut. The goal of the cut placement is to find a balance point that can eliminate all strut-strut intersections while preserving individual struts as much as possible. The output of this step is a modified lattice structure with no nodes, leaving only open struts.

Step 2 designs surface patches that can fill the openness introduced by Step 1. The goal is to design replacement nodal shapes that closely resemble the original ones. Given the non-linear nature of this shape design problem, a meta-heuristic optimizer is introduced to achieve optimized nodal shapes. Multiple optimization objectives are defined to ensure the nodal shapes are smooth and free from self-penetration. The output is a set of surface patches that approximate the original nodal shapes.

Step 3 stitches the surface patches from Step 2 with the open struts from Step 1 to construct a B-rep model of the lattice structure. With high-quality nodal shapes in place, this stitching process is straightforward and ensures that the final geometry is a valid solid model---free of self-penetrations, non-manifold elements, and cracks. The resulting model closely aligns with the original geometry of the lattice structure and consists of planar or curved faces (not limited to triangular faces as in prior methods~\cite{verma2020combinatorial, 2017_light-weight-triangulation}), allowing for more flexible and accurate lattice construction.

In the additional step of multiresolution triangular mesh generation, the constructed B-rep model serves as the base, and its surface patches are subdivided according to a specified error threshold. Given that the patches have simple topologies---either triangular or rectangular---ensuring consistent subdivision along shared edges is straightforward. Despite its simplicity, this approach enables the creation of adaptive triangles, offering a balance between computational efficiency and geometric accuracy.

\subsection{Optimal cutting}
\label{sec:optimal_cut}
Cutting struts can eliminate intersections, but excessive cutting may result in unnecessary losses, leaving little of the original structure intact. Therefore, the problem is to identify the optimal cut placements for each strut, which can remove all intersections while minimizing the amount of cutting. Without loss of generality, we can focus on a set of struts $\{e_i\}_{i=1}^n \subseteq E$ adjacent to the same node $v \in V$ and determine the minimum cutting length for each strut in $\{e_i\}$ required to eliminate the intersections, where the cutting lengths are measured from $v$ to the cut placements.

In the following, we begin with the simplest case, where two struts intersect at the same node, and demonstrate how to find the optimal cut placements. We then show that more general cases---multiple struts intersecting at the same node---can be cast as multiple applications of the simplest case, but executed in a carefully designed sequence.

\begin{figure}[htb]
    \centering
    \includegraphics[width=\linewidth]{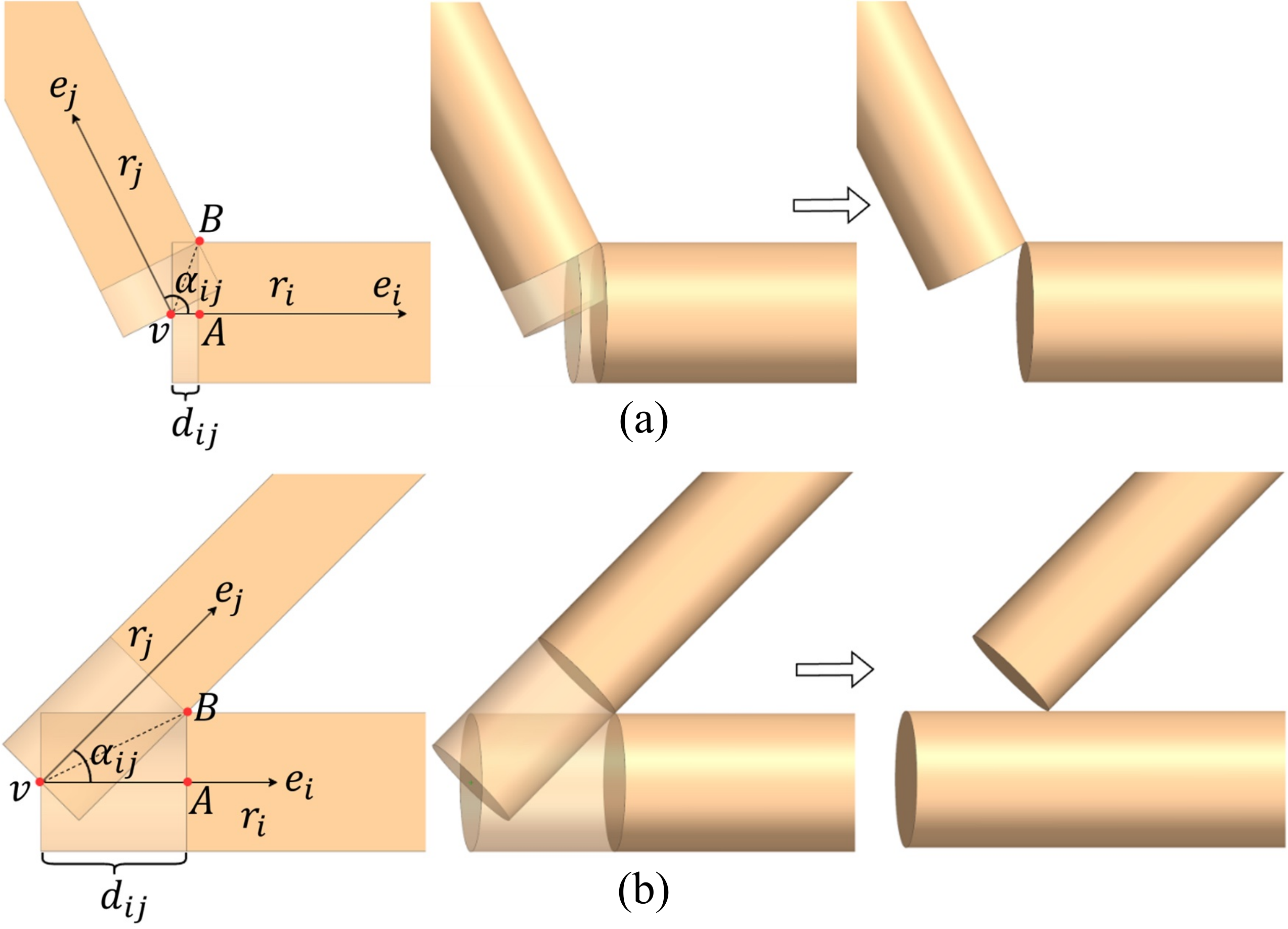}
    \caption{Minimum cutting length to avoid intersection with a pair of struts: (a) two struts with angle $> 90\degree$; and (b) two struts with angle $< 90\degree$}
    \label{fig:strut-strut-intersection} 
\end{figure}

Let $e_i$ and $e_j$ be two struts intersecting at node $v$, with radii $r_i$ and $r_j$, respectively. To simplify the notation, the (normalized) orientation vectors of the struts will also be denoted as $\boldsymbol{e_{i}}$ and $\boldsymbol{e_{j}}$, with $v$ as the starting point. Figure~\ref{fig:strut-strut-intersection} illustrates these two struts and their sectional view for deriving the optimal cutting length. Clearly, for this simple case, the intersection point $B$ is exactly where the cutting operation conduct, and the point $A$ is the cut placement for the corresponding strut $e_i$. Let $\alpha_{ij}$ the angle between $\boldsymbol{e_{i}}$ and $\boldsymbol{e_{j}}$, $\omega$ the angle between vector $\boldsymbol{vB}$ and $\boldsymbol{vA}$, the following equation holds true:

\begin{equation}
    |\boldsymbol{vB}|=\cfrac{r_{i}}{\sin\omega}=\cfrac{r_{j}}{\sin\left(\alpha_{ij}-\omega\right)}
\end{equation}
Subsequently, by employing the expansion of trigonometric functions, we can derive the following equation:

\begin{equation}
    \begin{aligned}
        & \frac{\sin\alpha_{ij}\cos\omega-\cos\alpha_{ij}\sin\omega}{\sin\omega}=\frac{r_{j}}{r_{i}} \\
 & \tan\omega=\frac{\sin\alpha_{ij}}{(r_{j}/r_{i})+\cos\alpha_{ij}}
    \end{aligned}
\end{equation}
Then the minimum cutting length $d_{ij}$ for the strut $e_i$ with respect to the strut $e_j$ to avoid intersection is given by:

\begin{equation}
    d_{ij} = |\boldsymbol{vA}| =\cfrac{r_i}{\tan\omega} =\cfrac{r_j+r_i\cos{\alpha_{ij}}}{\sin{\alpha_{ij}}}
\end{equation}
Note that when $\alpha_{ij}<90\degree$, only the strut with the smaller radius needs to be cut, as shown in Fig.~\ref{fig:strut-strut-intersection} (b).

\begin{figure*}[htb]
    \centering
    \includegraphics[width=\linewidth]{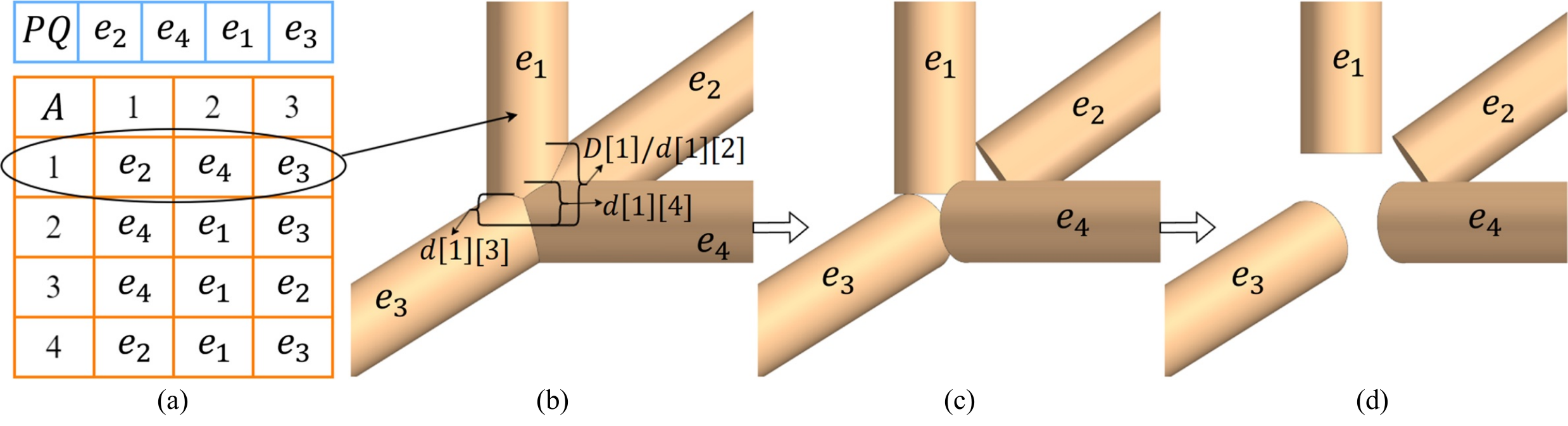}
    \caption{Optimal cutting: (a) data structure $PQ$ and $A$; (b) corresponding example of intersected struts; (c) result of optimal cutting; and (d) result of optimal cutting post-processing}
    \label{fig:optimal-cut}
\end{figure*}

Regarding the more general cases involving the intersection of multiple struts, above pair-wise computing of cutting lenths is evaluated for all struts 
connecting to the same node, and we maintain these lengths and their maximum value $D_i$ for each strut:
\begin{equation}
    D_i = \mathop{\max}_{j\in[1,n]
    \atop
    j \neq i} d_{ij}
    \label{eq:max-intersection-length}
\end{equation}
Obviously, the simplest way to avoid intersection for these struts is to cut them with their corresponding $D_i$, however, this can cause excessive cutting since the struts are interrelated. Consider the case of multiple struts shown in Fig. \ref{fig:optimal-cut} (b), in which the maximum $D_i$ among these struts is $D_2$. When strut $e_2$ is truncated by $D_2$, the remaining struts no longer need to be cut by their intersection placements with $e_2$, and similarly for the case where strut $e_4$ is trimmed by the updated $D_4$ which is computed pair-wise with strut $e_1$ and $e_3$. By iteratively cutting struts with the maximum $D_i$, we can make one strut free of intersection at a time without excessive cutting. Therefore, we propose to use the priority queue data structure to deal with the multiple struts cutting problem.

\begin{figure}[htb]
\includegraphics[width=\linewidth]{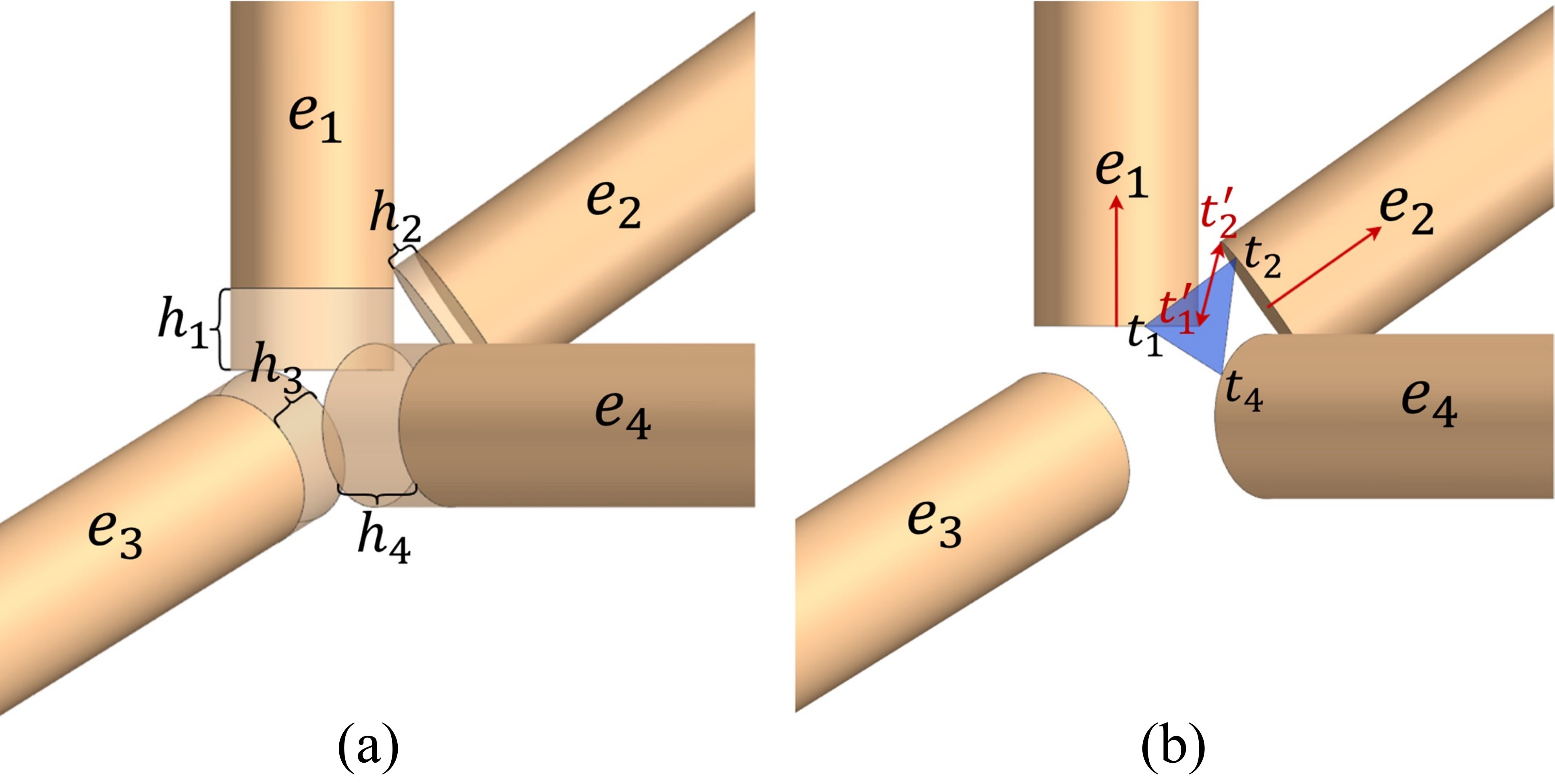}
\caption{Illustration of two objective functions for optimal cutting post-processing: (a) summation of new cutting lengths; (b) $|\boldsymbol{t_{1}^{\prime}t_{2}^{\prime}}|$ should be shorter to make node construction easier}
\label{fig:obj_fun} 
\end{figure}

\begin{algorithm}[htbp]
    \caption{Optimal cutting}
    \label{algo:optimalCut}
    \begin{algorithmic}[1]
    \raggedright
    \REQUIRE{A priority queue $PQ=\{e[i]\}_{i=1}^n$ of struts on one node organized by their maximum intersection lengths $\{D[i]\}_{i=1}^{n}$, an adjacency list $\{A[i][j]\}_{i=1,j=1}^{n,n-1}$ in which each row is organized as a priority queue by the intersection lengths $\{d[i][j]\}_{i=1,j=1}^{n,n-1}$ of neighboring struts on this strut $e[i]$}
  \ENSURE{An array of optimal cutting lengths $\{C[i]\}_{i=1}^{n}$ for struts on the same node}

  \STATE $C \leftarrow \{ 0 \}$
  \STATE $visited \leftarrow \{ \FALSE \}$
  \WHILE{$PQ \neq \emptyset$}
  \STATE $e[head] \leftarrow$ top($PQ$)
  \WHILE{$A[head] \neq \emptyset$}
  \STATE $e[neighbor] \leftarrow$ top($A[head]$)
  \IF{$visited[neighbor]$}
  \STATE pop($A[head]$)
  \IF{$A[head] \neq \emptyset$}
  \STATE continue
  \ELSE
  \STATE break
  \ENDIF
  \ENDIF
  
  \IF{$D[head] \neq d[head][neighbor]$}
  \STATE $D[head] \leftarrow d[head][neighbor]$
  \STATE updatePriorityQueue($PQ$)
  \STATE break
  \ENDIF
  \IF{$D[head] > C[head]$}
  \STATE $C[head] \leftarrow D[head]$
  \ENDIF
  \FOR{each strut $e[neighbor]$ in $A[head]$}
  \IF{$visited[neighbor] \neq \TRUE$}
  \STATE $L \leftarrow$ computeInterLength($e[neighbor]$,$e[head]$)
  
  \IF{$L > C[neighbor]$}
  \STATE $C[neighbor] \leftarrow L$
  \ENDIF
  \ENDIF
  \STATE pop($A[head]$)
  \ENDFOR
  \ENDWHILE
  \STATE $visited[head] \leftarrow \TRUE$
  \STATE pop($PQ$)
  \ENDWHILE
  \end{algorithmic}
\end{algorithm}

Specifically, given a sequence of struts $\{e_i\}_{i=1}^{n}$ intersecting at one node, we leverage a priority queue $PQ$ to maintain them. The priority queue is constructed according to the value $D_i$ of each strut, so that when the cutting operation is conducted, we always cut the strut with deepest intersection extent at first, this scheme will ensure the optimality. Furthermore, we utilize an adjacency list $\{A[i][j]\}_{i=1,j=1}^{n,n-1}$ to describe the intersection relationship among these struts, since each strut intersects with other $n-1$ struts  on the node, the element number of each $A[i]$ would be $n-1$. Note that strut elements in each row $A[i]$ are also organized as a priority queue by the intersection length $d_{ij}$ of neighboring strut $e_j$ on this strut $e_i$. The data structures and a simple example of struts are illustrated in Fig.~\ref{fig:optimal-cut} (a) and (b). The algorithm iteratively accesses the strut element in $PQ$ and then its neighboring struts in $A$ to update the optimal cutting length for each strut, and the strut elements are sequentially popped from $A$ and $PQ$ until they are both empty. For struts in Fig.~\ref{fig:optimal-cut} (b), the cutting process sequentially trims $e_2$, $e_4$, $e_1$ and $e_3$, as evidenced by the result in Fig.~\ref{fig:optimal-cut} (c). We give algorithm~\ref{algo:optimalCut} for more details about this cutting procedure. 

However, after above operation, struts are still huddled together which hinders the process of lattice node construction. We propose a post-processing of optimal cutting which uses meta-heuristic optimization to determine the further cut placements for the necessity of node building. We adopt the Grey Wolf Optimizer (GWO) algorithm as the choice of heuristic method and define corresponding objective functions for above mentioned purpose. As illustrated in Fig.~\ref{fig:obj_fun} (a), to maintain a controllable struts cutting, the first objective measures the summation of new cutting lengths which can be formulated as: 

\begin{equation}
    f_1=\sum_{i=1}^{n}{h_i}^2\\
\label{eq1}
\end{equation}
where $h_1, h_2, \cdots, h_n$ represent cutting lengths measured from the optimal cut placements to the newly updated cut positions for each strut.

As for the construction-orient objective function, its primary role is to determine which struts require further cutting. To this end, we turn our attention to the process of node surfaces generation. As depicted in Fig.~\ref{fig:pipeline}, to fill the openness of these four struts after optimal cutting, we first derive the projection points onto a unit sphere by projecting along struts' respective orientation vectors, and use a simple Delaunay triangulation to obtain topology among these points, then basic nodal triangles are acquired by connecting the end faces of four struts following the topology. One of these triangles which connects struts $e_1$, $e_2$, and $e_4$ is drawn in Fig.~\ref{fig:obj_fun} (b). This triangle is denoted by vertices $t_1$, $t_4$, and $t_2$ on each strut, and it's obvious from the figure that since the end face of strut $e_1$ is far away from the end face of strut $e_2$, the triangle edge $t_{1}t_{2}$ connecting these two struts is likely to intersect with strut $e_1$. To avoid triangle-strut intersection and inferior surface quality, $e_1$ should be further cut so as to make the end faces of $e_1$ and $e_2$ closer. Let $|\boldsymbol{t_{1}^{\prime}t_{2}^{\prime}}|$ to be the shortest line segment between end faces of $e_1$ and $e_2$, it's evident that $|\boldsymbol{t_{1}^{\prime}t_{2}^{\prime}}|$ should be shorter. Denote $\boldsymbol{e_{1}}$,$\boldsymbol{e_{2}}$ as normalized orientation vectors of these two struts, it can be observed that $\boldsymbol{t_{1}^{\prime}t_{2}^{\prime}}\cdot \boldsymbol{e_{1}} >0$ and $\boldsymbol{t_{2}^{\prime}t_{1}^{\prime}}\cdot \boldsymbol{e_{2}}<0$, we accumulate such $|\boldsymbol{t_{1}^{\prime}t_{2}^{\prime}}|^2$ to form the second objective function: 
\begin{equation}
\begin{split}
    & f_2=\sum_{\substack{i=1,j=1\\i\neq j}}^{n,n}{|\boldsymbol{t_{i}^{\prime}t_{j}^{\prime}}|}^2\\
    & if \{i,j\}\in\{Topology Set\}\&\& (\boldsymbol{t_{i}^{\prime}t_{j}^{\prime}}\cdot \boldsymbol{e_{i}} >0~||~\boldsymbol{t_{j}^{\prime}t_{i}^{\prime}}\cdot \boldsymbol{e_{j}}>0)
\end{split}
\end{equation}
with $\{TopologySet\}$ refers to the set of edge pairs of triangles that are generated from the Delaunay triangulation, the closest two points $\boldsymbol{t_{i}^{\prime}}$ and $\boldsymbol{t_{j}^{\prime}}$ on strut $e_i$ and $e_j$ are calculated by:
\begin{equation}
\begin{split}
    & \boldsymbol{t_{i}^{\prime}}=\cfrac{\boldsymbol{e_{i}} \times (\boldsymbol{e_{j}} \times \boldsymbol{e_{i}})}{|\boldsymbol{e_{i}} \times (\boldsymbol{e_{j}} \times \boldsymbol{e_{i}})|} \cdot r_i+\boldsymbol{e_{i}}\cdot(c_i+h_i)\\
    & \boldsymbol{t_{j}^{\prime}}=\cfrac{\boldsymbol{e_{j}} \times (\boldsymbol{e_{i}} \times \boldsymbol{e_{j}})}{|\boldsymbol{e_{j}} \times (\boldsymbol{e_{i}} \times \boldsymbol{e_{j}})|} \cdot r_j+\boldsymbol{e_{j}}\cdot(c_j+h_j)
\end{split}
\end{equation}
where operators ``$\times$'' and ``$\cdot$'' denote cross and dot product, respectively. $c_i$ and $c_j$ are the optimal cutting lengths of strut $e_i$ and $e_j$ output by algorithm~\ref{algo:optimalCut}.

Then a simple weighted addition makes up the final form of the objective function for the optimization process:
\begin{equation}
    F_{obj}=w_1f_1+w_2f_2
\end{equation}
where $w_1$ and $w_2$ are parameters that determine the weighted balance between the cutting length and the ease of node construction, and the result of optimal cutting post-processing is shown in Fig.~\ref{fig:optimal-cut} (d).

\begin{figure}[h]
\centering
\includegraphics[width=3.2in]{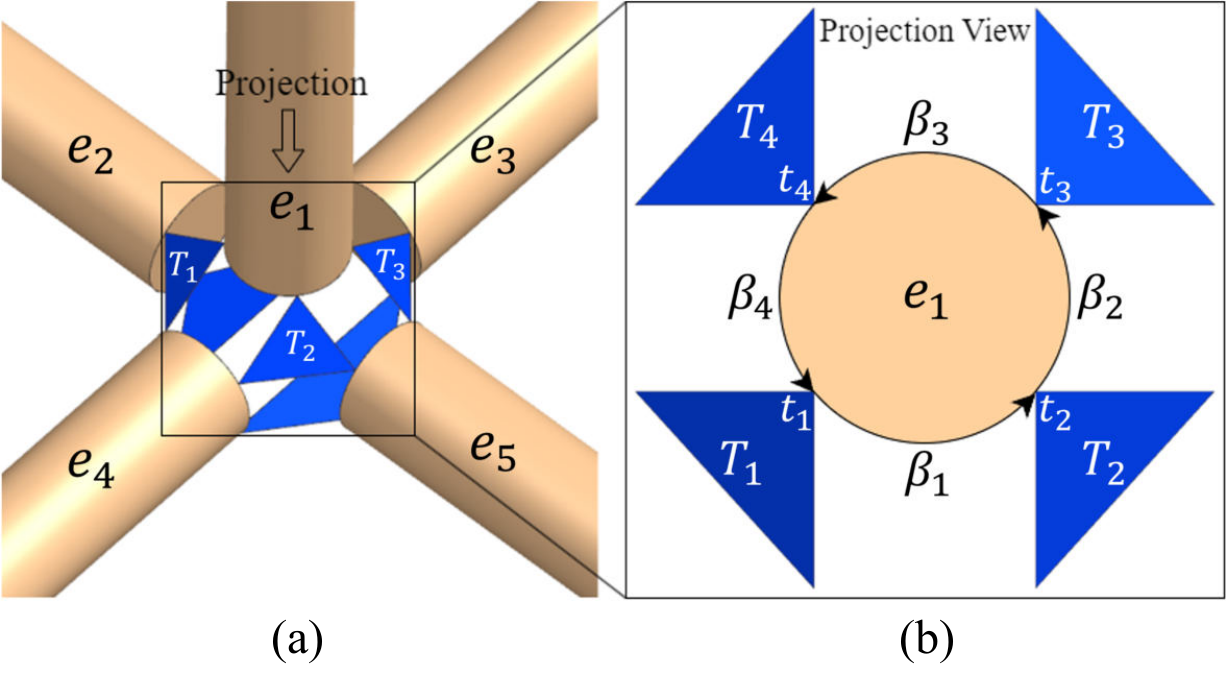}
\caption{Illustration of the nodal triangles adjustment: (a) nodal triangles on strut end faces; (b) projection view of the end face plane of strut $e_1$}
\label{adjustPoint} 
\end{figure}

\subsection{Optimal nodal shape design}
\label{sec:optimization}

\begin{figure}[h]
\centering
\includegraphics[width=3.3in]{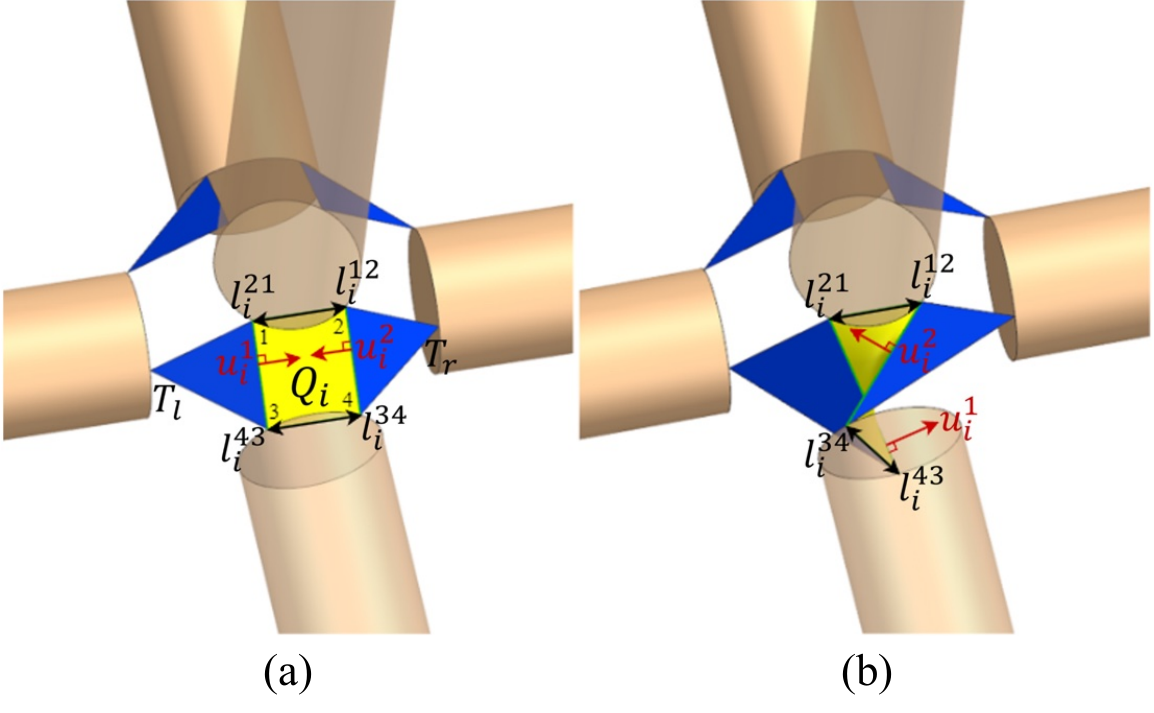}
\caption{Illustration of the first objective function for optimal nodal shape design: (a) regular quad face with smooth transition to adjacent triangles; (b) flipped quad face caused by overlapped triangles}
\label{F21} 
\end{figure}

\begin{figure}[htb]
\centering
\includegraphics[width=3.1in]{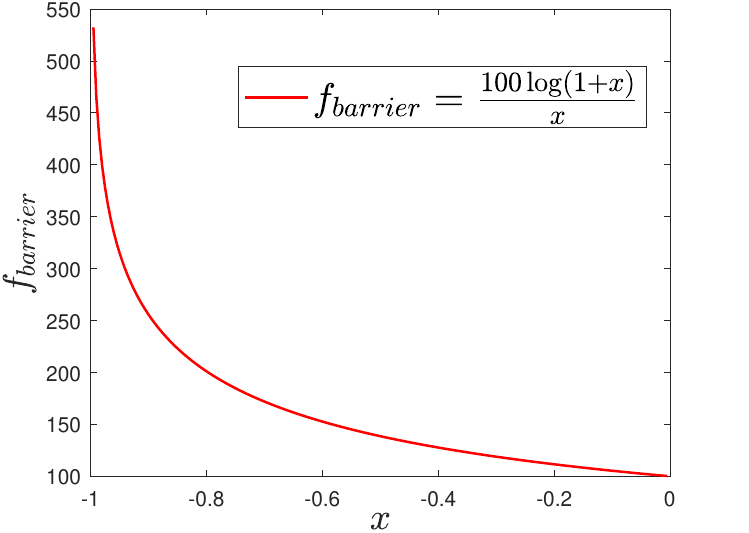}
\caption{Illustration of the adopted barrier function which is utilized to prevent corner cases}
\label{barrier} 
\end{figure}

\begin{figure}[h]
\includegraphics[width=3.2in]{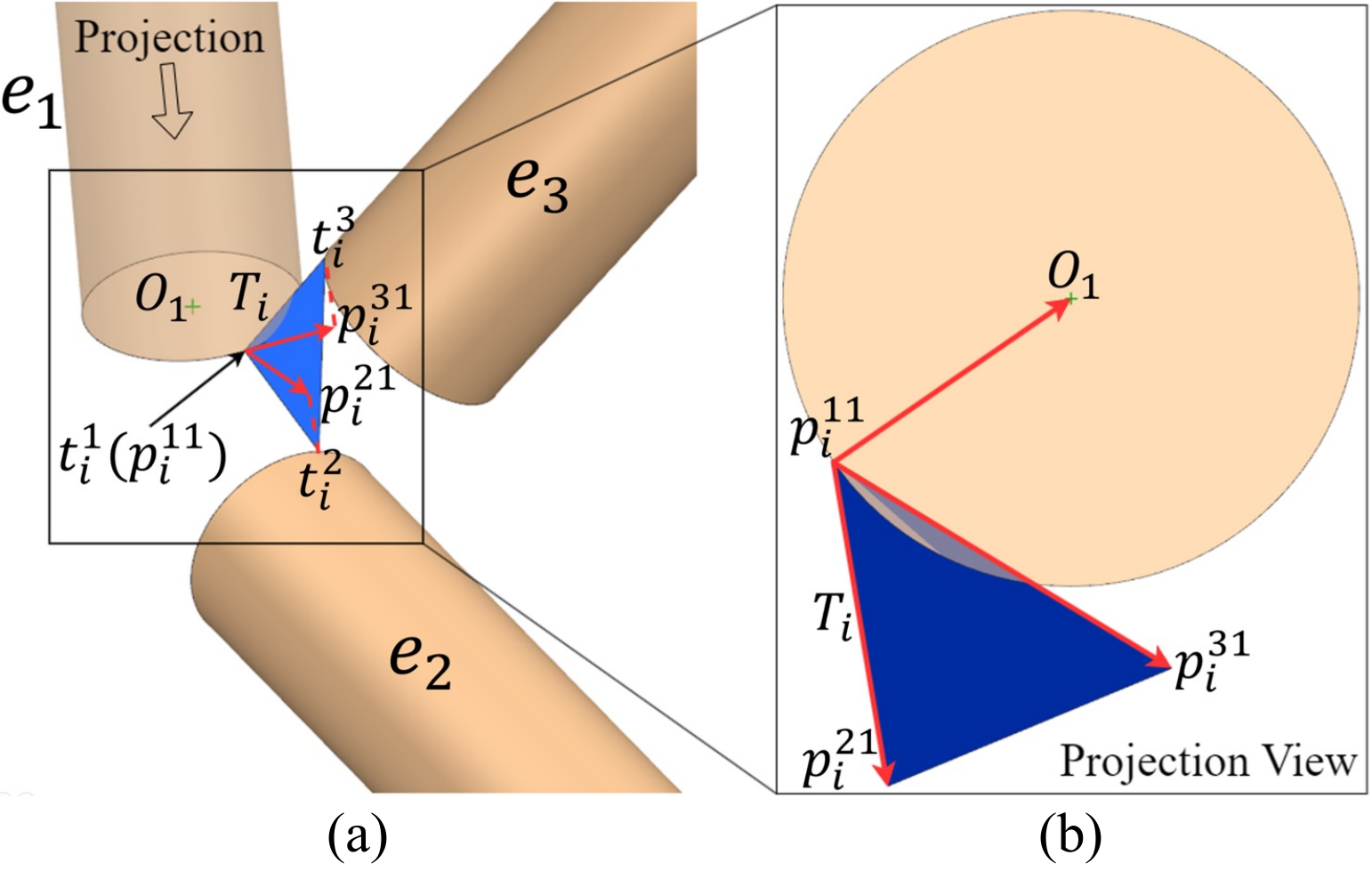}
\caption{Illustration of the second objective function for optimal nodal shape design: (a) strut $e_1$ intersects with triangle $\triangle t_i^1t_i^2t_i^3$; (b) projection view of the end face plane of strut $e_1$}
\label{F22} 
\end{figure}


To offer a quasi-optimal control of the lattice node shapes, this paper proposes to use optimization-based method to design the nodes. As shown in Fig.~\ref{adjustPoint}~(a), nodal triangles generated from the Delaunay triangulation are placed on corresponding strut end faces. From the projection view of strut $e_1$~(Fig.~\ref{adjustPoint}~(b)), 4 vertices of 4 triangles are placed and then adjusted along the circle of $e_1$, each vertex, $t_i$, is positioned by rotating the preceding vertex, $t_{i-1}$, counterclockwise by an angle of $\beta_{i-1}$ degrees about the center of the circle. The optimization algorithm will compute and update each $\beta_{i}$ with the constraint of $\beta_1+\beta_2+\beta_3+\beta_4=360\degree$. Through traversing every strut and evaluating these angles, the entire node shape is adjusted. With the feature of flexibility and simplicity, GWO is still utilized for this shape optimization task and two well-defined objective functions are introduced below.


Smoothness has been a significant criteria for surface quality evaluation. As depicted in Fig.~\ref{F21} (a), the first objective function aims to get a smooth transition between the quad face $Q_i$ and its two adjacent triangles, $T_l$ and $T_r$. Note that this quad face is bounded by two adjacent triangle edges and two straight lines connecting these two edges, and its 4 vertices are denoted by number 1-4. Since the quad face is generally not a plane, it's not easy to evaluate the dihedral angles between it and its two adjacent triangles. To solve this issue, we first compute two vectors $\boldsymbol{u_{i}^{1}}$, $\boldsymbol{u_{i}^{2}}$ which are perpendicular to adjacent triangle edges $\boldsymbol{l_{i}^{31}}$, $\boldsymbol{l_{i}^{24}}$, and meanwhile parallel to corresponding triangle planes $T_l$, $T_r$: 
\begin{equation}
    \begin{split}
        & \boldsymbol{u_{i}^{1}}=\boldsymbol{l_{i}^{31}} \times \boldsymbol{n_{l}}\\
    & \boldsymbol{u_{i}^{2}}=\boldsymbol{l_{i}^{24}} \times \boldsymbol{n_{r}}
    \end{split}
\end{equation}
where $\boldsymbol{n_{l}}$ and $\boldsymbol{n_{r}}$ are normalized normal vectors of $T_l$ and $T_r$. Let two straight lines connecting $T_l$ and $T_r$ to be two pairs of vectors $\boldsymbol{l_{i}^{12}}$, $\boldsymbol{l_{i}^{21}}$, $\boldsymbol{l_{i}^{34}}$, $\boldsymbol{l_{i}^{43}}$. Then we measure the angles between $\boldsymbol{u_{i}^{1}}$ and $\boldsymbol{l_{i}^{12}}$, $\boldsymbol{u_{i}^{1}}$ and $\boldsymbol{l_{i}^{34}}$, $\boldsymbol{u_{i}^{2}}$ and $\boldsymbol{l_{i}^{21}}$, $\boldsymbol{u_{i}^{2}}$ and $\boldsymbol{l_{i}^{43}}$. When $Q_i$ and $T_l$, $T_r$ lie in the same plane, with the edges $\boldsymbol{l_{i}^{31}}$, $\boldsymbol{l_{i}^{24}}$ aligned, the measures of the four angles mentioned above are zero, and resulting in a very regular quad face. 

Therefore, for each quad face $Q_i, i\in [1,N_Q]$, $N_Q$ the number of quad faces, let $\theta_{i1},\theta_{i2}$ the angles between $\boldsymbol{u_{i}^{1}}$ and $\boldsymbol{l_{i}^{12}}$, $\boldsymbol{l_{i}^{34}}$, $\theta_{i3},\theta_{i4}$ the angles between $\boldsymbol{u_{i}^{2}}$ and $\boldsymbol{l_{i}^{21}}$, $\boldsymbol{l_{i}^{43}}$, and use the cosine values of them to formulate the objective function: 
\begin{equation}
    f_1^{\prime}=\sum_{i=1}^{N_Q}\sum_{j=1}^{4}(\cos \theta_{ij}-1)^2
    \label{eq_6}
\end{equation}
which is an increment function, when this function value is decreasing, it indicates that the angles are approaching zero, resulting in a smoother transition between the quad face and the triangles. Note that the use of cosine functions here, rather than the angles themselves, is intended to mitigate the nonlinear effects that would otherwise be introduced by the application of inverse trigonometric functions.

Additionally, as shown in Fig.~\ref{F21} (b), when two nodal triangles overlap with each other, the quad face can flip over, leading to poor surface quality or even worse, a non-manifold geometry. We use the barrier function scheme~\cite{ames2019control} to penalize and avoid such corner cases. The function $f_{barrier}=w_b\log(1+x)/x, x \in (-1,0)$ is adopted for this purpose, when the variable $x$ decreases from 0, the function value grows exponentially, as illustrated in Fig.~\ref{barrier}. Note that when two or more of the four angles ($\theta_{i1},\dots,\theta_{i4}$)  are larger than $90\degree$, it's considered as a flipped case. In Fig.~\ref{F21} (b), for instance, $\theta_{i2}$ and $\theta_{i4}$ are both larger than $90\degree$, the cosine of these angles is a decrement function ranging from -1 to 0 with respect to the angle, substitute it into the barrier function to get a big penalty value:

\begin{equation}
    f_{b1} = \left\{\begin{array}{rcl}
      \sum_{i=1}^{N_Q}\sum_{j=1}^4 \cfrac{\log(1+\cos{\theta_{ij}})}{\cos{\theta_{ij}}} & if~flipped~\&\&~\theta_{ij}>90\degree \\
         0 & otherwise
     \end{array}\right . 
\end{equation}


The second objective function aims to avoid intersection between triangles and struts. As shown in Fig.~\ref{F22}, strut $e_1$ intersects with triangle $T_i$ which is represented by vertices $\boldsymbol{t_{i}^{1}}$, $\boldsymbol{t_{i}^{2}}$, and $\boldsymbol{t_{i}^{3}}$. This circumstance not only generates distorted surfaces, but also introduces self-penetration of nodes, resulting in an invalid lattice model. To tackle this challenge, we project vertices of $T_i$ onto the end face plane of $e_1$, and projection points $\boldsymbol{p_{i}^{11}}$, $\boldsymbol{p_{i}^{21}}$, $\boldsymbol{p_{i}^{31}}$ are computed by:
\begin{equation}
    \begin{split}
        & \boldsymbol{p_{i}^{11}} = \boldsymbol{t_{i}^{1}}\\
        & \boldsymbol{p_{i}^{21}} = \boldsymbol{t_{i}^{2}}-(\boldsymbol{t_{i}^{2}}-\boldsymbol{t_{i}^{1}}) \cdot \boldsymbol{e_{1}} \cdot \boldsymbol{e_{1}}\\
        & \boldsymbol{p_{i}^{31}} = \boldsymbol{t_{i}^{3}}-(\boldsymbol{t_{i}^{3}}-\boldsymbol{t_{i}^{1}}) \cdot \boldsymbol{e_{1}} \cdot \boldsymbol{e_{1}}
    \end{split}
\end{equation}
where the notation $\boldsymbol{p_{i}^{jk}} (j,k\in[1,3])$ represents the projection point of $\boldsymbol{t_{i}^{j}}$ on the end face plane of strut $e_k$. Given the center $\boldsymbol{O_{k}}(k\in[1,3])$ of the circular end face, it can be observed that $\angle O_1p_i^{11}p_i^{31} < 90\degree$ and $\angle O_1p_i^{11}p_i^{21} > 90\degree$, a preferred position for $\boldsymbol{p_{i}^{11}}$ should be rotating it counterclockwise about $\boldsymbol{O_{1}}$ to make above two angles both larger than $90\degree$. 

With the consideration of reducing nonlinear factors, we still use the cosine value of these angles to construct the objective function. Specifically, for each nodal triangle $T_i$, $i \in [1,N_T]$, $N_T$ the number of nodal triangles, let $\varphi_{i1}=\angle O_1p_i^{11}p_i^{21}, \varphi_{i2}=\angle O_1p_i^{11}p_i^{31}$, and repeat above projection computing for the other two struts $e_2, e_3$ to get the rest four angles:
\begin{equation}
    \begin{split}
        & \varphi_{i3}=\angle O_2p_i^{22}p_i^{12}\\
        & \varphi_{i4}=\angle O_2p_i^{22}p_i^{32}\\
        & \varphi_{i5}=\angle O_3p_i^{33}p_i^{13}\\
        & \varphi_{i6}=\angle O_3p_i^{33}p_i^{23}
    \end{split}
\end{equation}
Through iteratively processing nodal triangles, the second objective is illustrated as:
\begin{equation}
    f_2^{\prime}=\sum_{i=1}^{N_T}\sum_{j=1}^{6}\cos{\varphi_{ij}}
\end{equation}
so that when the $\cos{\varphi_{ij}}$ gradually descends to a negative value, the corresponding triangle is adjusted to be placed at the right position.

Meanwhile, we introduce the barrier function to penalize the triangle-strut intersection cases. It can be derived from Sec.~\ref{sec:optimal_cut} that when $(\boldsymbol{e_{1}}\cdot \boldsymbol{t_{i}^{1}t_{i}^{2}}>0~\&\&~\varphi_{i1}<90\degree)~||~(\boldsymbol{e_{1}}\cdot \boldsymbol{t_{i}^{1}t_{i}^{3}}>0~\&\&~\varphi_{i2}<90\degree)$, strut $e_1$ would intersect with the triangle $T_i$. Substitute corresponding positive $\cos{\varphi_{ij}} \in (0,1]$ into the barrier function to get a large penalty value:
\begin{equation}
    f_{b2}=\left\{\begin{array}{rcl}
        \sum_{i=1}^{N_T}\sum_{j=1}^6 \cfrac{\log(1-\cos{\varphi_{ij}})}{-\cos{\varphi_{ij}}} & if~inter.~\&\&~\varphi_{ij}<90\degree \\
        0 & otherwise
    \end{array}\right .     
\end{equation}


At last, a simple weighted addition makes up the final form of the objective function for the optimal nodal shape design process:
\begin{equation}
    F_{obj}^{\prime}=w_1^{\prime}f_1^{\prime}+w_2^{\prime}f_2^{\prime}+w_{b1}f_{b1}+w_{b2}f_{b2}
\end{equation}
where the parameters $w_1^{\prime},w_2^{\prime},w_{b1},w_{b2}$ are carefully adjusted and tested to ensure both high node quality and fast convergence.

\subsection{Lattice stitching}
\label{sec:B-rep}
\begin{figure}[h]
\centering
\includegraphics[width=2.8in]{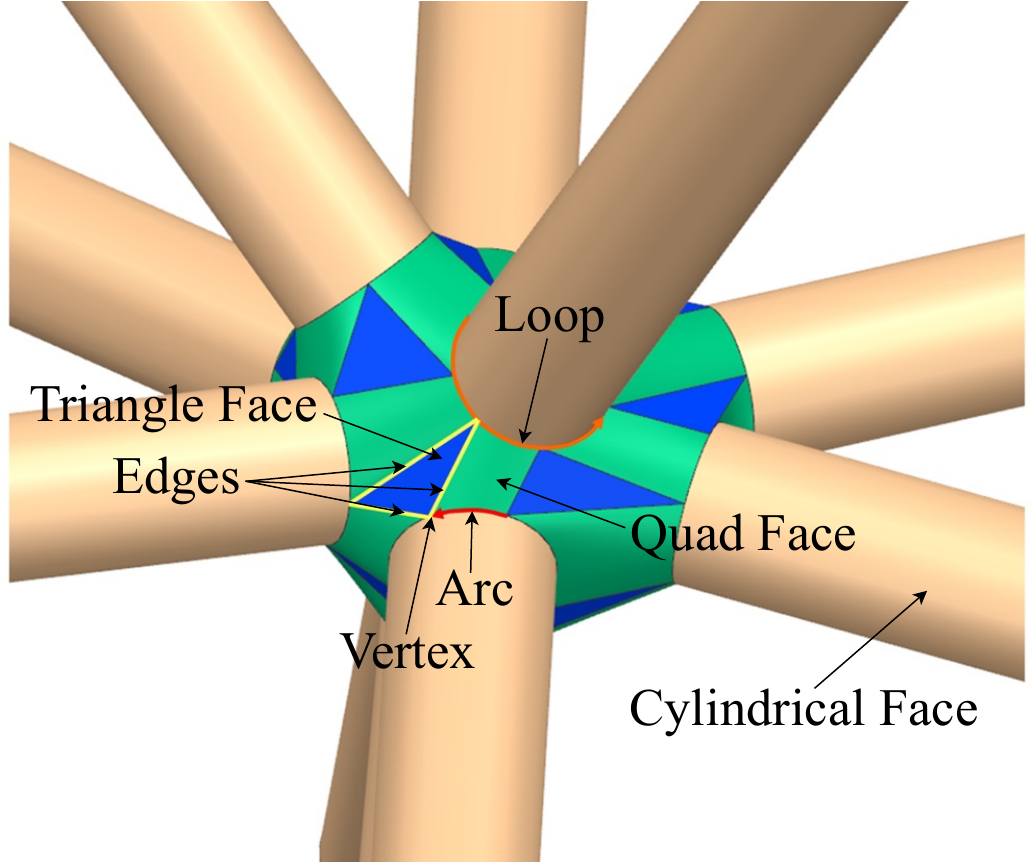}
\caption{Illustration of B-rep elements on a lattice cell}
\label{BRep} 
\end{figure}


With nodal triangles and corresponding surface topology in place, the approximated nodes are constructed easily and the B-rep model of lattice structures can be obtained by stitching the approximated portions back into the lattice. The lattice structure consists of cylindrical struts and designed central nodes. For simplicity of illustration, consider each central node equipped with connecting struts as a lattice cell, and the whole structure as the assemble of these cells. Then each lattice cell is the region bounded by a set of faces bounded by a set of edges meeting in a set of vertices.

We thus represent the elements (faces, arcs, loops, edges, vertices) (Fig.~\ref{BRep}) of the boundary of the cell as follows:
\begin{itemize}[label=\textbullet,font=\large,labelsep=-4pt]
    \item \quad Face: Faces of lattice cell can be categorized into three types. The first is triangle face which is the optimization object bounded by edges, the second is quad face which is represented by two triangle edges and two circular arcs, the last is cylindrical face which is constructed by two circle loops. Given the struts number $n$, the number of triangle faces and quad faces are calculated by $N_T=2n-4$ and $N_Q=3n-6$, respectively. 
    \item \quad Arc: Each arc of the cell is both a portion of a quad face and a loop. When the optimization of nodal triangles is done, the start and end point of each arc are determined by these triangles and their adjacency relationship. With radius of the strut it belongs to, the radian angle and length of the arc can be easily calculated.
    \item \quad Loop: A loop is a cycle of arcs, it can be represented by simply referring to its arcs and connecting them end to end. Each loop is oriented such that when walking upright along it, the quad face it bounds lies on its right.
    \item \quad Edge: Each edge is defined by two triangle vertices with six coordinates.
    \item \quad Vertex: Each vertex of the cell lies on the intersection of four faces, two of which are quad faces and the other two are triangle face and cylindrical face, the entire node shape changes by movement of these vertices.
\end{itemize}


\subsection{Multiresolution triangular mesh generation}
\begin{figure}[t]
\centering
\includegraphics[width=2.8in]{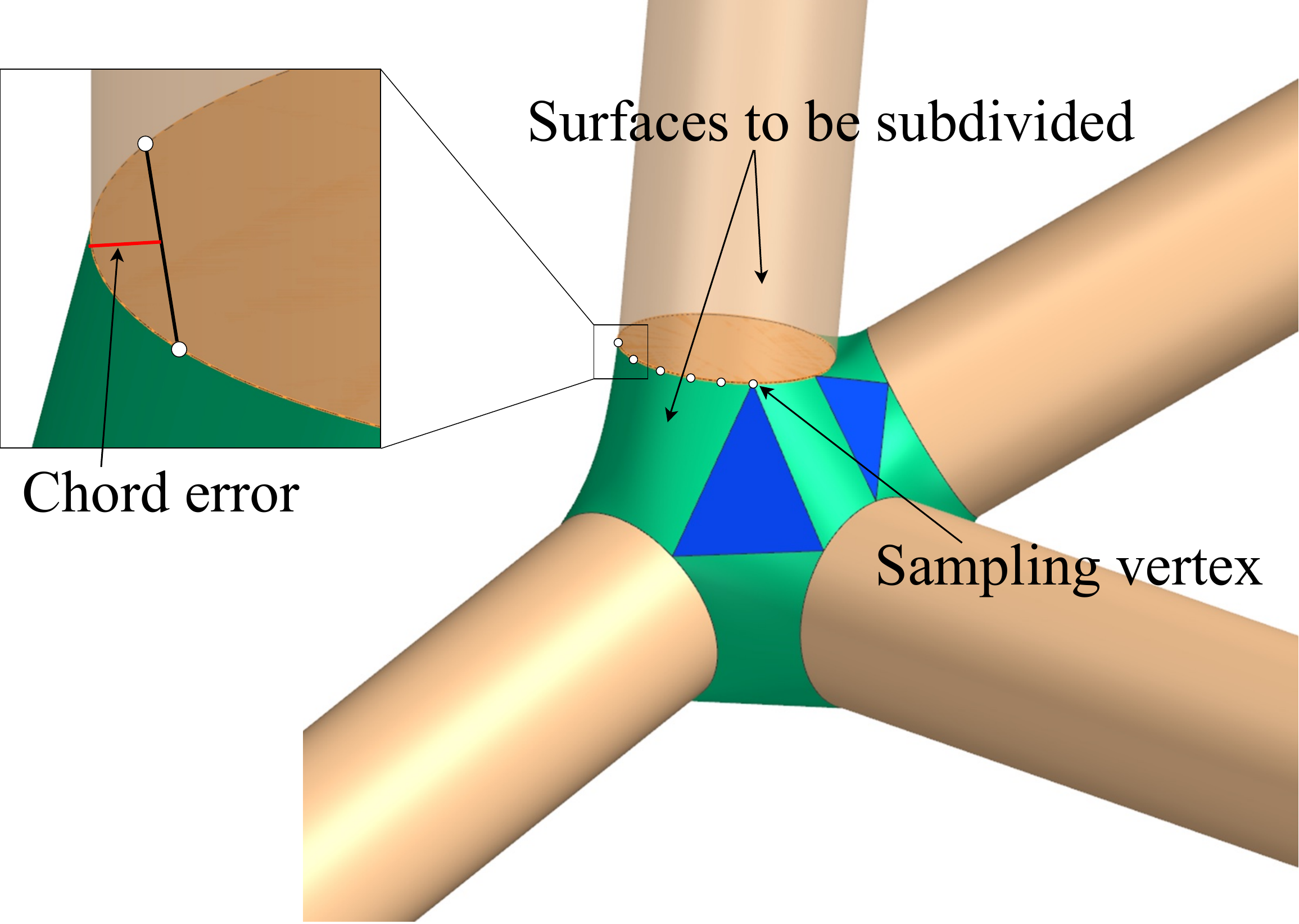}
\caption{Illustration of subdividing arcs according to the chord error} 
\label{fig:mesh} 
\end{figure}
Based on the constructed B-rep model,  triangulating lattice structures can be simply done by subdividing the model's arcs. To obtain a high-quality triangulation,
we make subdivided vertices uniformly distributed on the arc and implement the subdivision process with controllable chord error, as shown in Fig.~\ref{fig:mesh}. If we want to reuse the B-rep model to generate a triangulation at a different resolution, we only need to re-subdivide the arcs according to the new chord error.

Assuming the chord error (measured in terms of the percentage of the radius) is $CE$, and the parameter range of the arc is from $a_1$ to $a_2$, the number of the subdivided vertices $N$ is given by:
\begin{equation}
    N=\left\lfloor \cfrac{a_2-a_1}{2\cos^{-1}(1-CE)} \right\rfloor +1
\end{equation}
where $\lfloor\rfloor$ is a floor function to get the integer part of the value. The formula finds the minimum number of vertices, which satisfy the specified chord error. In practical applications, multiple triangulation results are required, like visualization. With uniform subdivision, we can define the parameters $s_i$ on the original arcs by:
\begin{equation}
    s_i=\cfrac{a_2-a_1}{N} \cdot i +a_1\quad i=0,1,\ldots,N
\end{equation}
Using the above equation, all arcs on the two arc loops of a strut and two arcs of a quad face can be subdivided. Then a cylindrical surface and a curved quad surface can be easily triangulated by connecting these vertices.

In summary, taking advantage of optimal cutting, our method can offer a closer lattice shape approximation compared with \cite{wu2020chocc}, and through conducting optimal nodal shape design, our method gives a more robust solution for lattice node construction compared with \cite{verma2020combinatorial}. In the meantime, our method generates lattice model consists of planar and curved faces, enabling a more flexible and accurate lattice structure construction.



\section{Results and Discussion}
\label{sec:results}


To verify effectiveness and test performance of the proposed method,  a Nvidia
RTX 3060 Ti GPU paired with an Intel Core i7-12700F CPU were utilized for the test. All experiments were conducted using C++ 17 and CUDA 11.8, operating on Windows 10 system.


Based on this implementation, 10 case studies are to be presented to demonstrate the effectiveness of the proposed method. Cases 1-9 considered 9 lattice structures which are categorized into three different complex levels, and each of them visualized the lattice B-rep models and their triangulation results. Case 10 showed the results of constructing lattice of different strut radii. Shape deviation analysis (Table~\ref{tab:1} and Fig.~\ref{comparison}) were used to carry out comparisons with
other two approximation methods: Combinatorial method~\cite{verma2020combinatorial} and CHoCC method~\cite{wu2020chocc}, more importantly, to verify whether the challenges were resolved. Performance analysis (Table~\ref{tab:2} and Fig.~\ref{runTime}) tested converging speed of our method and the flexibility of the method to be applied to GPU parallelism. 
\subsection{Examples}

\begin{figure*}[!ht]
\centering
\includegraphics[width=6.8in]{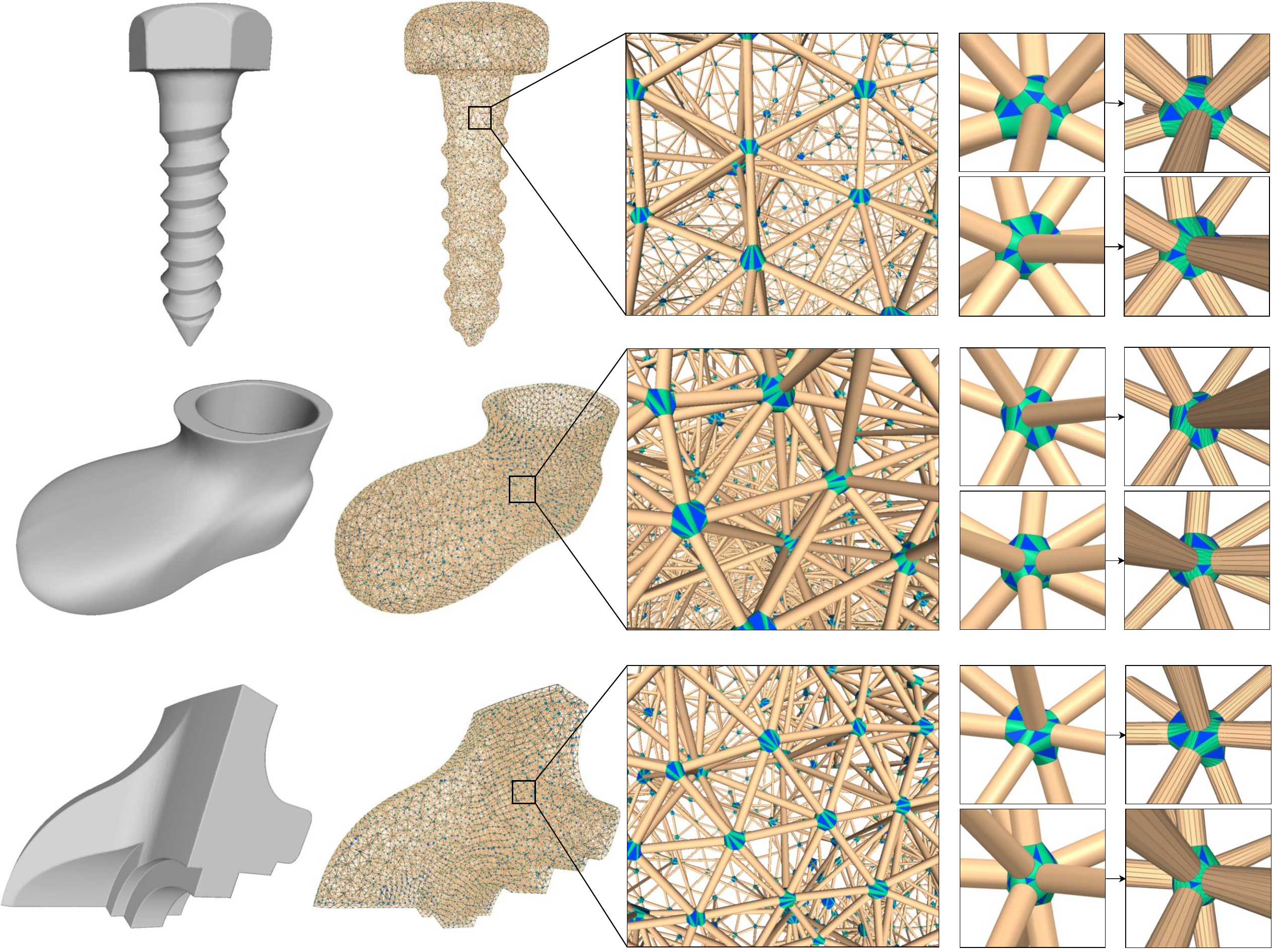}
\caption{Original models, B-rep models, and corresponding triangulation results of simple lattice structures, from top to bottom are case studies of the screw, boot, and bracket models, respectively}
\label{simple_case}
\end{figure*}

\begin{figure*}[!ht]
\centering
\includegraphics[width=6.8in]{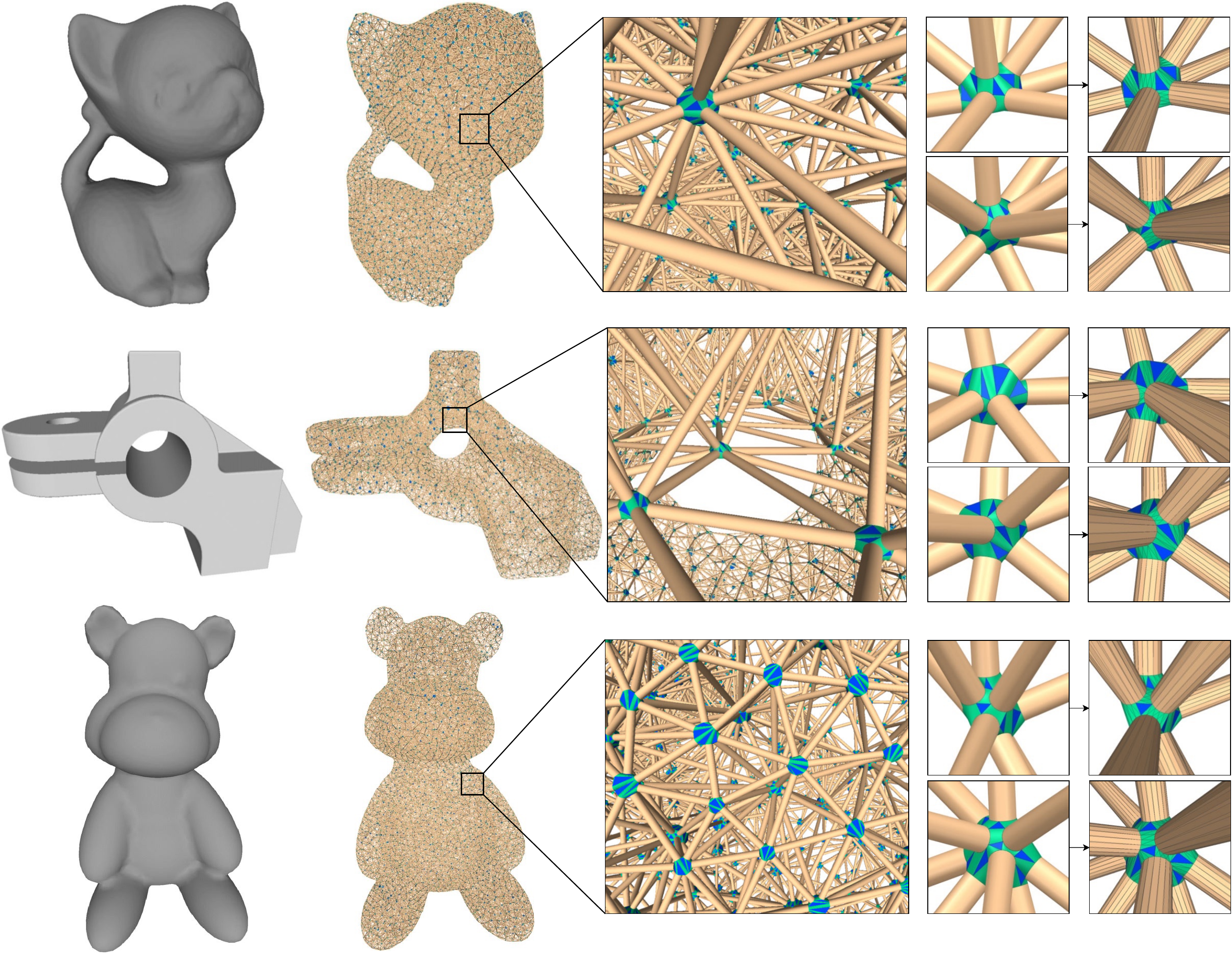}
\caption{Original models, B-rep models, and corresponding triangulation results of medium complex lattice structures, from top to bottom are case studies of the kitten, hinge, and teddy models, respectively}
\label{medium_case}
\end{figure*}

\begin{figure*}[!ht]
\centering
\includegraphics[width=6.8in]{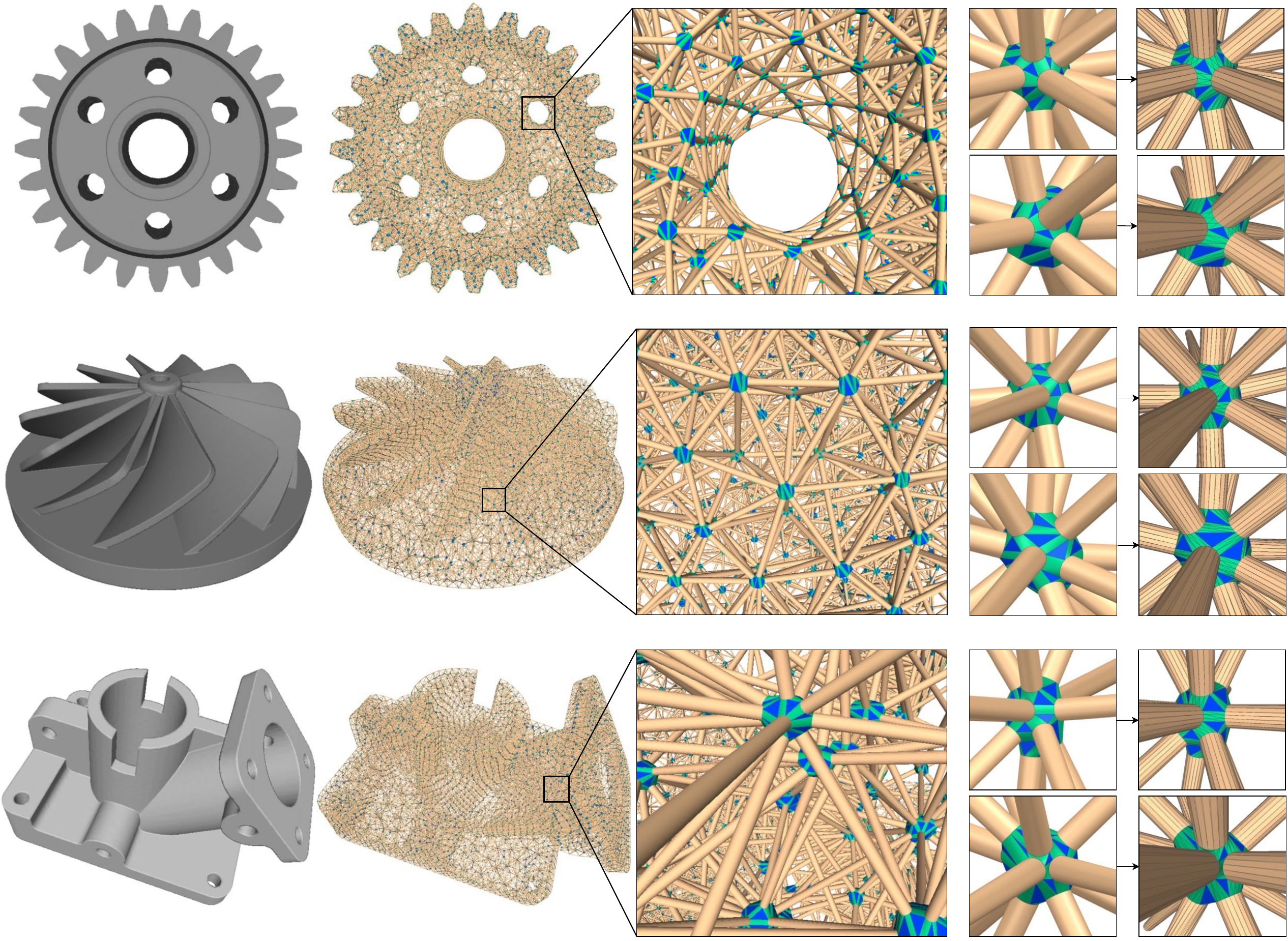}
\caption{Original models, B-rep models, and corresponding triangulation results of complex lattice structures, from top to bottom are case studies of the gear, impeller, and connector models, respectively}
\label{complex_case}
\end{figure*}

\begin{figure*}[!ht]
\centering
\includegraphics[width=6.8in]{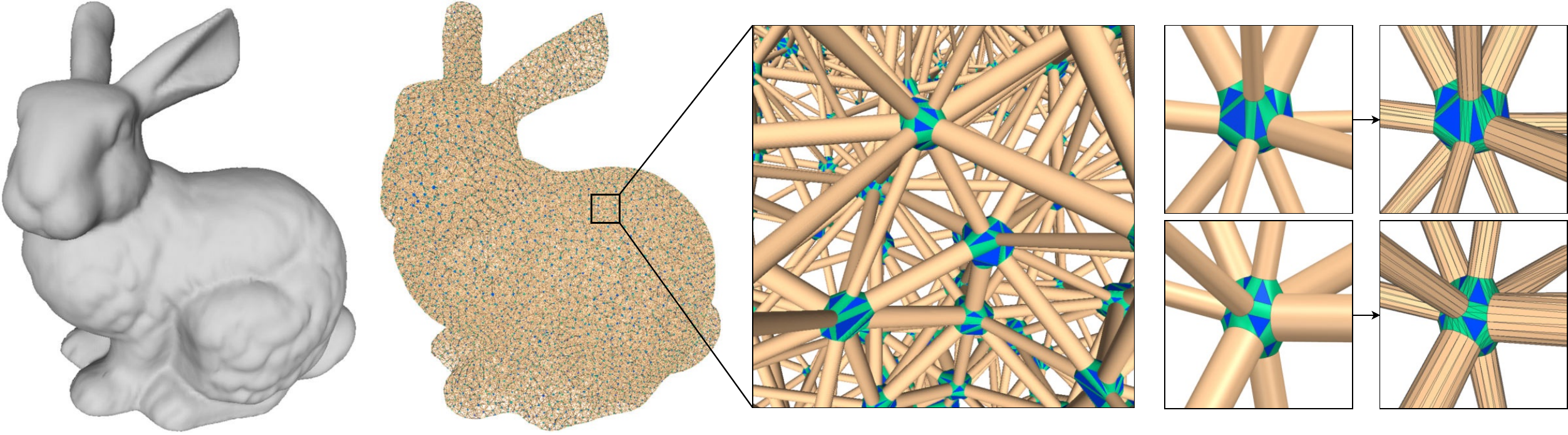}
\caption{Illustration of lattice structures with different strut radii}
\label{diffRadius}
\end{figure*}

\begin{table*}[!ht]
    \centering
    \setlength\extrarowheight{3pt}
    \setlength{\abovecaptionskip}{0cm}
    \footnotesize
    \caption{Comparisons of our
method with other two approximation methods for shape deviation}
    \setlength{\tabcolsep}{3.5mm} {
    \begin{tabular}{c c c c c c}
    \hline
        \multirow{2}{*}{Models} & \multirow{2}{*}{Methods} & \multicolumn{3}{c}{Deviation} & {Improvements}\\
        \cline{3-5}
        & &  Min & Max & Avg & (w.r.t. Average deviation)\\
        \hline
            \multirow{3}{*}{Screw} & Our method & 31.74\% & 136.17\% & 93.34\% \\
             & Combinatorial & 72.9\% & 136.17\% & 109.18\% & 14.51\% \\
             & CHoCC & 115.27\%  & 136.17\% & 124.63\% & 25.1\% \\
             \hline
             \multirow{3}{*}{Boot} & Our method & 34.58\% & 139.41\% & 95.59\% \\
             & Combinatorial & 76.33\% & 139.41\% & 111.4\% & 14.19\% \\
             & CHoCC & 116.89\%  & 139.41\% & 127.83\% & 25.22\% \\
             \hline
             \multirow{3}{*}{Kitten} & Our method & 45.37\% & 160.28\% & 101.82\% \\
             & Combinatorial & 85.19\% & 160.28\% & 124.21\% & 18.03\% \\
             & CHoCC & 121.4\%  & 160.28\% & 143.95\% & 29.27\% \\
             \hline
             \multirow{3}{*}{Teddy} & Our method & 42.31\% & 156.05\% & 100.75\% \\
             & Combinatorial & 80.72\% & 156.05\% & 121.33\% & 16.96\% \\
             & CHoCC & 123.37\%  & 156.05\% & 141.8\% & 28.95\% \\
             \hline
             \multirow{3}{*}{Gear} & Our method & 49.29\% & 181.43\% & 106.39\% \\
             & Combinatorial & 88.64\% & 181.43\% & 133.27\% & 20.17\% \\
             & CHoCC & 123.21\%  & 181.43\% & 154.75\% & 31.25\% \\
             \hline
             \multirow{3}{*}{Impeller} & Our method & 50.8\% & 185.97\% & 107.64\% \\
             & Combinatorial & 91.15\% & 185.97\% & 136.12\% & 20.92\% \\
             & CHoCC & 124.73\%  & 185.97\% & 159.18\% & 32.38\% \\
        \hline
    \end{tabular} }
    \label{tab:1}
\end{table*}

\begin{table}[!ht]
    \centering
    \setlength\extrarowheight{3pt}
    \setlength{\abovecaptionskip}{0cm}
    \footnotesize
    \caption{The iteration number stats of phase \ding{172} optimal cutting post-processing and phase \ding{173} optimal nodal shape design for different models}
    \setlength{\tabcolsep}{2mm} {
    \begin{tabular}{l l l l l l l l}
    \hline
         \multirow{2}{*}{Model} & \multirow{2}{*}{Degree} & \multicolumn{3}{c}{Iteration number of \ding{172}} & \multicolumn{3}{c}{Iteration number of \ding{173}} \\
         \cline{3-8}
                    & &  Min & Max & Average & Min & Max & Average \\
        \hline
        Screw & 5-10 & 5 & 27 & 14 & 4 & 65 & 16\\
        Boot & 5-10 & 6 & 26 & 13 & 4 & 69 & 18\\
        Bracket & 5-10 & 5  & 26 & 13 & 6 & 62 & 16\\
        Kitten & 9-15 & 5 & 30 & 17 & 6 & 78 & 23\\
        Hinge & 9-15 & 7 & 35 & 19 & 6 & 85 & 26\\
        Teddy & 9-15 & 6 & 35 & 17 & 5 & 71 & 24\\
        Gear & 12-20 & 7 & 43 & 21 & 7 & 87 & 31\\
        Impeller & 12-20 & 8 & 49 & 23 & 5 & 90 & 35\\
        Connector & 12-20 & 8 & 48 & 22 & 7 & 85 & 33\\
    \hline
    \end{tabular} }
    \label{tab:2}
\end{table}

\begin{figure}[!ht]
\centering
\includegraphics[width=3.2in]{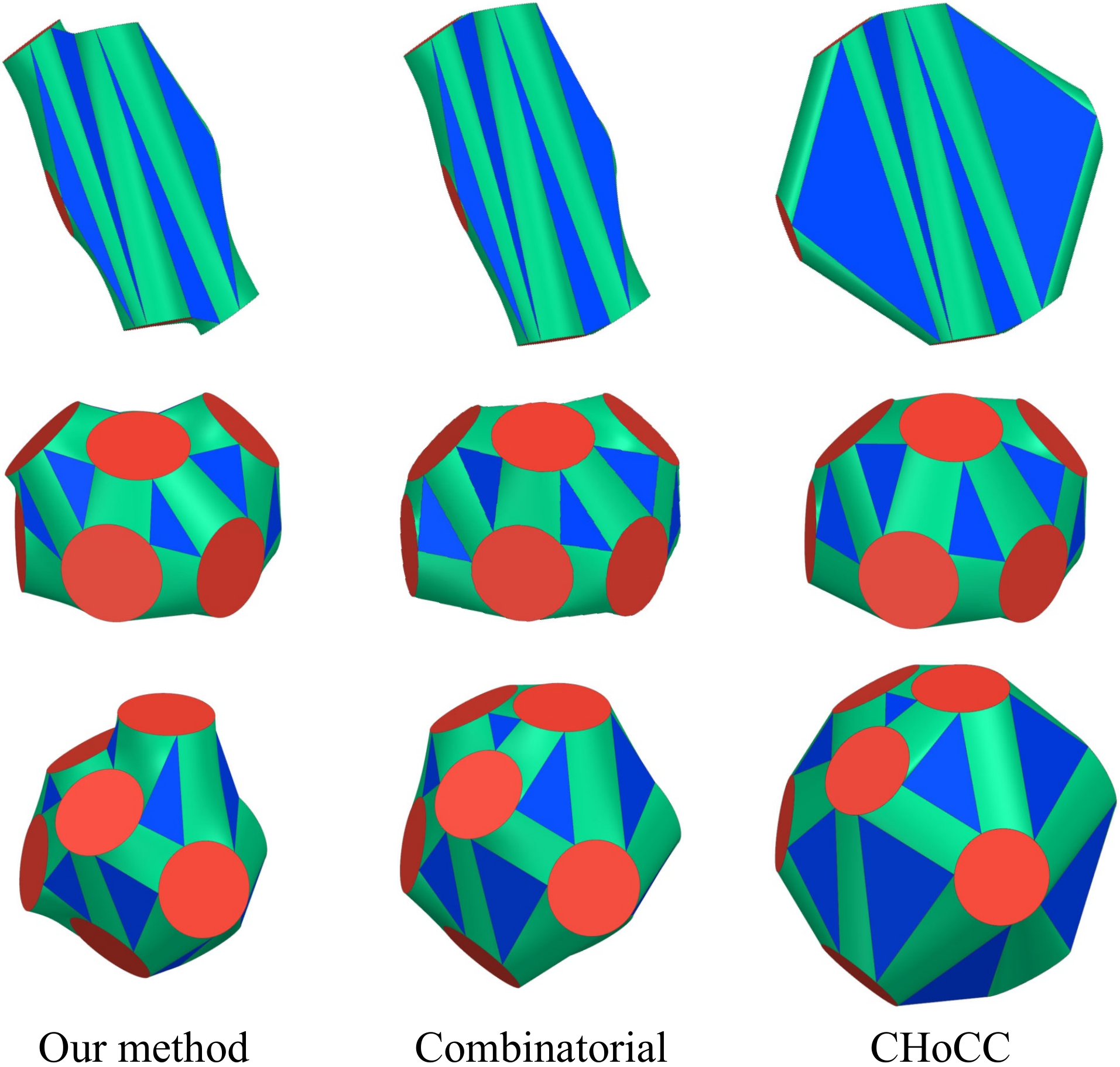}
\caption{Node shape comparison between our method and other two approximation methods, each row corresponds to the constructed nodes with the same configuration of struts}
\label{comparison} 
\end{figure}

\begin{figure}[!ht]
\centering
\includegraphics[width=3.3in]{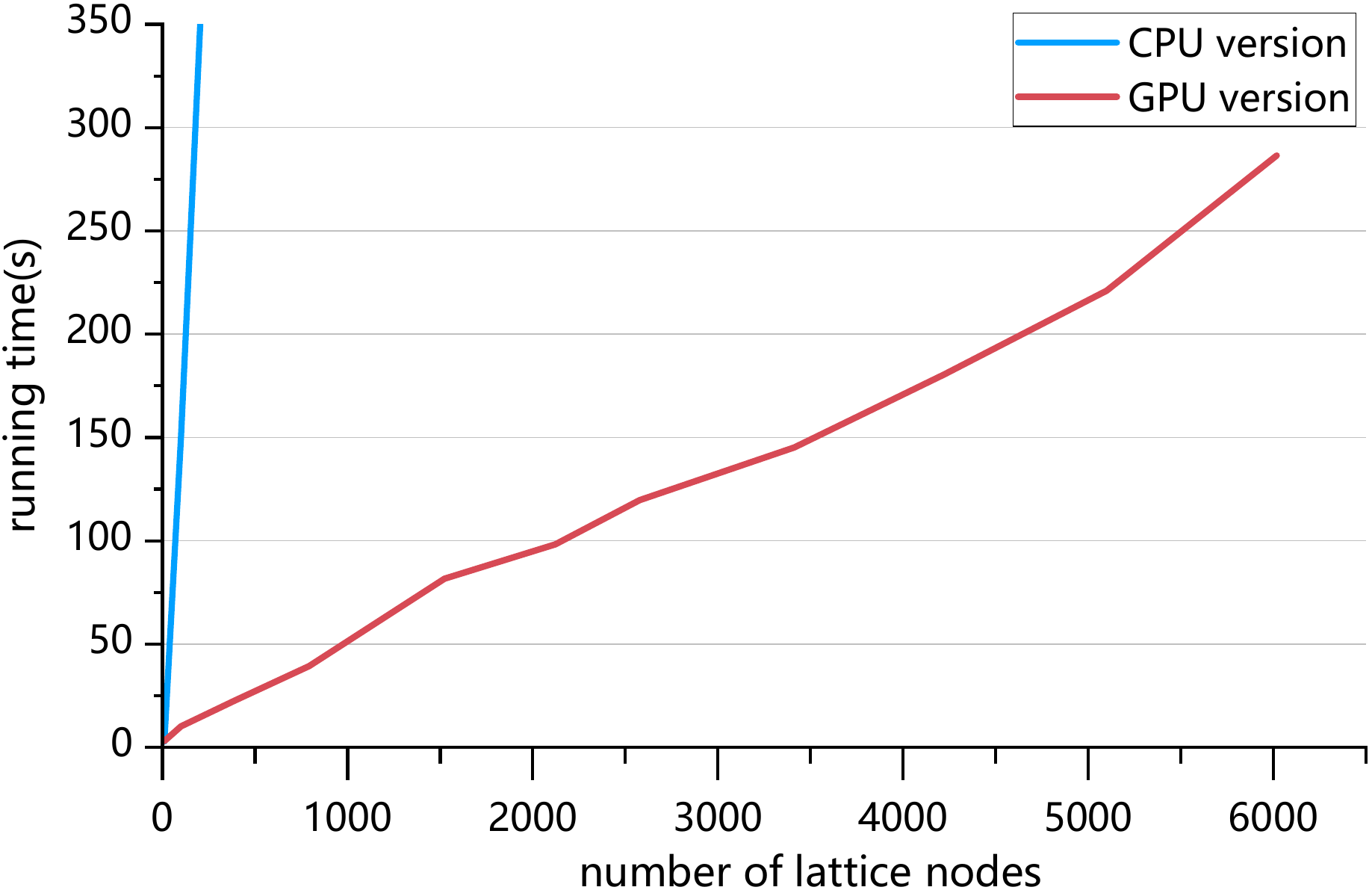}
\caption{Time used in constructing lattice structures with varying nodes number}
\label{runTime} 
\end{figure}


Case studies 1-3 (Fig.~\ref{simple_case}) considered three lattice structures with simple geometric shapes, the original model of them are typical solids without holes or complicated surfaces, and the degree of their nodes is about 5-10. Case studies 4-6 (Fig.~\ref{medium_case}) involved three medium complex lattice structures in which their original models consist of freeform surfaces, and the node degree ranged from 9 to 15. Case studies 7-9 (Fig.~\ref{complex_case}) analyzed three lattice structures of the complex level in which the original models contain multiple holes and intricate surface topology, some struts connect with narrow angles and the node degree mainly ranged from 12 to 20. Case study 10 (Fig.~\ref{diffRadius}) considered lattice structures with struts of different radii. Note that the triangulation operation for all of the case studies is also optimization-based to maintain uniform shapes of generated triangles.


The shape deviation analysis (Table~\ref{tab:1}) considered the relative minimum, maximum, and average shape deviation between original spherical nodes and simplified nodes constructed by different methods, in which the deviation is measured by computing the distance between approximated node surfaces and the spherical surface. Note that the original spherical node is a regular sphere with the radius equal to the maximum radius of the connecting struts.
Given a surface point $p_i$ on the constructed node corresponding to the lattice graph vertex $v_j \in V$, the original sphere radius $R$, the deviation is simply calculated by $||\boldsymbol{p_{i}v_{j}}|-R|/R$.
Furthermore, the visualization of several typical lattice nodes constructed by our method and other two approximation methods is given (Fig.~\ref{comparison}). From the statistics and the visualization, our method achieved the minimum shape deviation and obtained smooth node surfaces, which verify the effectiveness of the proposed approach.


The performance analysis first considered the efficiency of our method. As illustrated in Table~\ref{tab:2}, we tested iteration number of the optimization algorithm for 9 models. It can be observed that the GWO optimizer converged to a valid state within a few dozen iterations. Secondly, we analyzed the flexibility of our method to be applied to GPU parallelism. As shown in Fig.~\ref{runTime}, we compared running time of the optimization process for models of different nodes number. From the results, the GPU version of our method can handle lattice structures of varying scale within a reasonable time.


\subsection{Discussion and Limitations}

In all of the above lattice structures, the B-rep models are generated by the optimal cutting, optimal nodal shape design, and lattice stitching process. From Fig.~\ref{simple_case}, Fig.~\ref{medium_case}, Fig.~\ref{complex_case}, the sophisticated connections among struts are well represented while the lattice nodes are also constructed in satisfactory quality. In Fig.~\ref{diffRadius}, the modeling results demonstrate the capability of our method to construct lattice structures with struts of varying radii. From these case studies, the B-rep models of lattice nodes and struts are well evaluated, which validates the robustness of our method. 


From the stats in Table~\ref{tab:1}, the proposed method shows both lower minimum shape deviation and lower average shape deviation compared with the other two methods, particularly the CHoCC method, attributed to the optimal cutting. As depicted in Fig.~\ref{comparison}, compared with other two approximation methods, our method generated lattice nodes with reduced volume cost and facilitated a more natural connection among the struts. These stats and visualization results confirm that the proposed method provides better deviation control, less volume waste, and achieves superior surface quality. 


As shown in Table~\ref{tab:2}, during the optimal cutting post-processing phase, the average iteration count for all models are below 25, with the maximum iteration count not exceeding 50. In the optimal nodal shape design phase, the average number of iterations for these models are under 35, with the maximum iteration count remaining below 90, this demonstrates the validity of the objective functions' formulation. And in Fig.~\ref{runTime}, although the number of nodes and struts in the model increases significantly, the overall running time does not grow substantially compared with CPU version of the algorithm. For the lattice structure with 6018 nodes and 30578 struts, our method is able to complete the optimization within 5 minutes, which validates the advantages of our method being adapted for GPU parallelism.


In this paper, while various types of lattice structures are considered, certain corner cases may still pose challenges for the proposed method. Specifically, due to the struts cutting operation, when the length-to-diameter ratio of the struts are too small or when the struts intersect at very narrow angles, the calculated cutting lengths may exceed the strut lengths, which is not acceptable. However, these corner cases occur in less than 1\% of the case studies and can be easily addressed by either removing the affected struts or editing the input lattice graph.

\section{Conclusion}
\label{sec:conclusion}
This paper presents a novel method for constructing lattice structures, featuring a three-step process to design an optimal lattice shape that eliminates surface-surface intersections while preserving the original lattice geometry as much as possible. This approach not only resolves robustness issues in lattice structure construction but also minimizes shape deviations from the ideal lattice (whereas existing methods often fail to account for such deviations). This is achieved by integrating meta-heuristic optimization techniques into lattice structure design and defining multiple optimization objectives to ensure that the resulting shapes are smooth, free from self-penetration, and yield a valid solid B-rep model. Extensive case studies demonstrated that the method can provide a notable improvement over existing methods.

While the proposed method has demonstrated effectiveness in the case studies conducted, there is still room for improvement. Most notably, the current approach focuses exclusively on lattice struts with straight axes and profiles. Although these are the most commonly used lattice types, other non-linear structures, such as Quador~\cite{gupta2018quador}, are also of practical significance. Extending the method to accommodate a broader range of lattice structures is among the research studies to be carried out in the future. 

Moreover, the proposed optimization framework focuses primarily on geometric quality and topological validity in the construction of lattice structures, without considering their mechanical properties. For instance, smooth transitions between struts and nodal shapes can help reduce stress concentrations and improve ease of fabrication. Incorporating mechanical properties, such as stress distribution and material behavior, into the optimization process presents a promising opportunity for enhancing both the structural performance and manufacturability of lattice structures.


\section*{Acknowledgment} 
This work has been funded by the “Pioneer” and “Leading Goose” R\&D Program of Zhejiang Province (No. 2024C01103), the National Natural Science Foundation of China (No. 62102355), Natural Science Foundation of Zhejiang Province (No. LQ22F020012), and the Fundamental Research Funds for the Zhejiang Provincial Universities (No. 2023QZJH32).




\appendix   





 \bibliographystyle{elsarticle-num} 

\bibliography{refs} 



\end{document}